%% file: rtc.tex
\documentclass[10pt]{article}

\usepackage{grffile}
\usepackage{ctable}

\usepackage{amssymb}
\usepackage{amsmath}
\usepackage{mathrsfs}
\usepackage[dvips]{epsfig}
\usepackage[footnotesize]{caption}
\usepackage{graphicx}
\usepackage[all]{xy}
\usepackage{color}
\usepackage{multirow}
\usepackage{amsmath,amsfonts,amsthm,amssymb}
\usepackage{graphicx,import}
\usepackage{wrapfig}
\usepackage{float}
\usepackage{mathtools}
\usepackage{subcaption}
\usepackage{url}

\newtheorem{Lemma}{Lemma}[section]
\newtheorem{Theorem}{Theorem}
\newtheorem{Proposition}[Lemma]{Proposition}
\newtheorem{Corollary}[Lemma]{Corollary}

\newtheorem{Remark}[Lemma]{Remark}

\newenvironment{Proof}%
 {\begin{trivlist} \item[]{\bf Proof. }}%
 {\hspace*{\fill}$\rule{.4\baselineskip}{.4\baselineskip}$\end{trivlist}}

 {\begin{trivlist}\item[]\textbf{Acknowledgments.}}{\end{trivlist}}

\makeatletter\@addtoreset{figure}{section}\makeatother

\makeatletter \@addtoreset{equation}{section} \makeatother

\newcommand{\R}{\mathbb{R}}

\def\Re{\mathop{\mathrm{Re}}}
\def\Im{\mathop{\mathrm{Im}}}

\newcommand{\rmO}{\mathrm{O}}

\newcommand{\rmd}{\mathrm{d}}
\newcommand{\rme}{\mathrm{e}}
\newcommand{\rmi}{\mathrm{i}}

\newcommand{\sech}{\,\mathrm{sech}\,}
\newcommand{\eps}{\varepsilon}

\makeatletter
\newsavebox{\@brx}
\newcommand{\llangle}[1][]{\savebox{\@brx}{\(\m@th{#1\langle}\)}%
  \mathopen{\copy\@brx\kern-0.5\wd\@brx\usebox{\@brx}}}
\newcommand{\rrangle}[1][]{\savebox{\@brx}{\(\m@th{#1\rangle}\)}%
  \mathclose{\copy\@brx\kern-0.5\wd\@brx\usebox{\@brx}}}
\makeatother

\definecolor{Green}{rgb}{0.,0.4,0.}

\renewcommand{\leq}{\leqslant}
\renewcommand{\geq}{\geqslant}

\newcommand{\Rmnum}[1]{\uppercase\expandafter{\romannumeral #1\relax}}

\def\XXint#1#2#3{{\setbox0=\hbox{$#1{#2#3}{\int}$}
     \vcenter{\hbox{$#2#3$}}\kern-.5\wd0}}

\newfam\bifam
\font\tenbi=cmmib10 scaled \magstep1 \font\sevenbi=cmmib10 at 11pt
\font\fivebi=cmmib10 at 6pt \textfont\bifam = \tenbi
\scriptfont\bifam = \sevenbi \scriptscriptfont\bifam= \fivebi

\ifx\pdfoutput\undefined
   \pdffalse
\else
   \pdfoutput=1
   \pdftrue
\fi
\ifpdf
   \usepackage{graphicx}
   \usepackage{epstopdf}
\else
   \usepackage{graphicx}
\fi

\begin{document}

\begin{center}
{\fontsize{15}{15}\fontfamily{cmr}\fontseries{b}\selectfont    
Self-organized Clusters in Diffusive Run-and-Tumble Processes
}
\\[0.2in]
Patrick Flynn$\,^1$, Quinton Neville$\,^2$, Arnd Scheel$\,^3$\\
\textit{\footnotesize $\,^1$ Oregon State University, Department of Mathematics, Kidder Hall 368, Corvallis, OR 97331, USA\\
\textit{\footnotesize $\,^2$ St. Olaf College, Department of Mathematics, 1500 St. Olaf Ave., Northfield, MN 55057, USA}}\\
\textit{\footnotesize $\,^3$University of Minnesota, School of Mathematics,   206 Church St. S.E., Minneapolis, MN 55455, USA}\\

\date{\small \today} 
\end{center}

\begin{abstract}
\noindent 
We analyze a simplistic model for run-and-tumble dynamics, motivated by observations of complex spatio-temporal patterns in colonies of myxobacteria. In our model, agents run with fixed speed either left or right, and agents turn with a density-dependent nonlinear turning rate, in addition to diffusive Brownian motion. We show how a very simple nonlinearity in the turning rate can mediate the formation of self-organized stationary clusters and fronts. Phenomenologically, we demonstrate the formation of barriers, where high concentrations of agents at the boundary of a cluster, moving towards the center of a cluster, prevent the agents caught in the cluster from escaping. Mathematically, we analyze stationary solutions in a four-dimensional ODE with a conserved quantity and a reversibility symmetry, using a combination of bifurcation methods, geometric arguments, and numerical continuation. We also present numerical results on the temporal stability of the solutions found here. 
\end{abstract}
 \section{Introduction}
Consider two populations of agents with densities $u(t,x)$ and $v(t,x)$ on the real line $x\in\R$, diffusing with rate $\eps^2$, and moving with fixed speed to the left and to the right, respectively, In addition, we suppose that left- and right-moving agents interact through a tumbling mechanism, where left-moving agents tumble and become right-moving agents with rate $r(u,v)$. Assuming reflection symmetry, right-moving agents then tumble with rate $r(v,u)$, leading to the system of partial differential equations
\begin{equation}\begin{aligned}
u_t &= \varepsilon^2 u_{xx} + u_x - r(u,v) + r(v,u),\\
v_t &= \varepsilon^2 v_{xx} - v_x + r(u,v) - r(v,u).
\label{myxo_pde}
\end{aligned}
\end{equation}
Models of this form clearly represent a variety of ``traffic flow'' situations. Our interest is particular motivated by the formation of rippling patterns and fruiting bodies in colonies of myxobacteria \cite{lutscher2002emerging,Reichenbach65,SagerKaiser94,scheel2016wavenumber,ShimketsKaiser82,WelchKaiser01}. Indeed, tumbling dependent on encounters with other agents has been identified as a driver in the collective behavior of myxobacteria, which are capable of communicating upon end-to-end contact through the so-called C-signal \cite{SagerKaiser94}. We are interested here in fairly simple tumbling rates $r(u,v)$, that exhibit an increase depending on the concentration. We therefore consider monotone rates 
\begin{equation}\label{e:r}
 r(u,v)=u\cdot g(v),\qquad g(v)= \mu + \frac{v^p}{1 + \gamma v^q},
\end{equation}
where $\mu$ models spontaneous tumbling, $p$ a power law growth in the tumbling rate depending on head-to-head encounters, $\gamma$ a saturation level for the increase, and $p-q$ the power-law growth (or decay) of tumbling rates for high frequencies of head-to-head encounters.

It turns out that the analysis of stationary structure is somewhat intricate in this case and we therefore also study a somewhat flawed, simplistic example,
\begin{equation} \label{e:rs}
 r(u,v)=u\cdot g(u+v),\qquad g(w)=(w-1)(\gamma w-1),
\end{equation}
that is, tumbling rates depend on all encounters with  all agents, not only with agents traveling in the opposite direction. We refer to this latter case as an \emph{aggregate sensing}, and to the former case as \emph{head-on sensing}. Clearly, turning rates that combine both features, for instance $g=\mu+v^p/(1+(u+v)^q)$ with increase through head-on encounters and saturation from overcrowding, may well be more realistic, but we restrict ourselves here to the two cases \eqref{e:r} and \eqref{e:rs}, illustrating the somewhat universal scenario of cluster formation and growth throughout the family of possible nonlinear tumbling rates. 

More precisely, our interest here is in the possibility of stationary solutions to \eqref{e:r} or \eqref{e:rs} that possess limits
\[
 u(x)\to u_\pm,\qquad v(x)\to v_\pm,\qquad u'(x)\to 0,\qquad v'(x)\to 0, \qquad \text{for } x\to\pm\infty.
\]
We refer to the case $u_-=u_+,v_-=v_+$ as homoclinic structures, which turn out to be either clusters or gaps in the population density, and other cases as cluster boundaries. We also restrict to the case $u_\pm=v_\pm$, that is agents are asymptotically equidistributed across left- and right-moving populations. 

Our main results can be informally summarized as follows. 
\begin{enumerate}
\item There exist two one-parameter families of homoclinic structures, parameterized by the level of the background concentration $u_\pm=v_\pm$, one family representing a cluster and the other a gap in concentration. Both families limit on a constant density state, where the amplitude of the cluster or gap converges to zero, and both limit on the same pair of cluster boundaries, when the width of cluster or gap tends to infinity. 
\item  Direct simulations and numerical computation of spectra of the linearized operators suggest that clusters and gaps are stable for intermediate values of mass densities, and cluster boundaries are stable for sufficiently small values of the saturation $\gamma$. 
\end{enumerate}

\paragraph{Outline.}
We present analytical results on existence of homoclinic and heteroclinic orbits in Section \ref{section:timeindp}. Section \ref{section:numerics} contains numerical analysis of homoclinic and heteroclinic families, and Section \ref{section:direct} illustrates the results in direct simulations, complemented with a numerical stability analysis. 

\paragraph{Acknowledgment.} Most of this work was carried out during an REU project on ``Complex Systems'' at the University of Minnesota, funded through NSF grant  DMS-
1311740.

\section{Homoclinics, heteroclinics, clusters, and gaps}
\label{section:timeindp}
We analyze the stationary equations corresponding to \eqref{myxo_pde}. We first discuss the formulation of the equation for stationary solutions as a dynamical system in the spatial variable $x$, together with some qualitative properties of the resulting dynamical system, Section \ref{section:symmetry}. We then study the case of aggregate sensing  where  properties of stationary solutions are accessible ``explicitly'' in Section \ref{section:simplified}.  
In Section \ref{section:asymmetric}, we provide an analogous set of results for the case of head-on sensing, when explicit solutions are not available. 

\subsection{Spatial Dynamics and Symmetry}\label{section:symmetry}
We first consider general turning rates $r(u,v)$.  The symmetry of the system suggests $\rho = u+v$ and $m = u-v$ as a natural coordinate system. The quantity $r(u,v) - r(v,u)$ changes sign under the switching of $u$ and $v$, and therefore is odd in $m$.  Thus, as long as $r(u,v)-r(v,u)$ is smooth, there exists a smooth function $R(\rho,m)$ such that $mR(\rho,m) = -2(r(u,v) - r(v,u))$.  In $(\rho,m)$-coordinates, we find
\begin{equation}
\begin{aligned}
\rho_t &= \varepsilon^2 \rho_{xx} + m_x,\\
m_t &= \varepsilon^2 m_{xx} + \rho_x +mR(\rho,m).
\end{aligned}
\label{rho_m}
\end{equation}
% We write our equation as a system of three first order equations, with the linear component written in matrix form:
% \begin{equation}
% \frac{d}{dx}\begin{pmatrix} \rho \\ m \\ m_1 \end{pmatrix} = \begin{pmatrix} \theta/\varepsilon^2 \\ 0 \\ 0 \end{pmatrix} + \begin{pmatrix} 0 & -1/\varepsilon^2 & 0\\ -1/\varepsilon^2 & 0 & 1/\varepsilon^2 \\ 0 & 0 & 0 \end{pmatrix} \begin{pmatrix}\rho \\ m \\ m_1  \end{pmatrix} + \begin{pmatrix} 0 \\0\\-R(\rho,\rho')/\varepsilon^2\end{pmatrix}.javascript:void(0);
% \label{rho_only}
% \end{equation}
% Written in this form, it becomes clear that
We consider stationary solutions, that is, we set $\rho_t = m_t = 0$, thus obtaining a pair of second-order differential equations
\begin{equation}
\begin{aligned}
0&= \varepsilon^2 \rho'' + m',\\
0 &= \varepsilon^2 m'' + \rho' +mR(\rho,m).
\end{aligned}
\label{rho_m_0}
\end{equation}
This system possesses a first integral, $\theta :=\varepsilon^2 \rho' + m$, that is, $\theta$ is constant along any solution of \eqref{rho_m_0}.
We can rewrite the equation solely in terms of $\rho$, a transformation which in particular puts the  linear part in Jordan normal form,
\begin{equation}
\frac{d}{dx}\begin{pmatrix} \rho \\ \rho_1 \\ \rho_2 \end{pmatrix} =  \begin{pmatrix} 0 & 1 & 0\\ 0 & 0 & 1\\ 0 & 0 & 0 \end{pmatrix} \begin{pmatrix}\rho \\ \rho_1 \\ \rho_2  \end{pmatrix} + \begin{pmatrix} 0 \\0\\ \eps^{-4}\rho_1 + (\eps^{-4}\theta-\eps^{-2}\rho_1) R(\rho,\theta-\varepsilon^2\rho_1))/\varepsilon^2\end{pmatrix}.
\label{rho_system}
\end{equation}
This system is equivalent to the third-order ODE 
\begin{equation}
-\eps^4\rho'''+ \rho'+(\theta-\eps^2\rho')R(\rho,\theta-\varepsilon^2\rho')=0.
\label{little_rho_system}
\end{equation}
We shall focus on solutions asymptotic to symmetric states, where $u=v$, or $m=0$. We therefore choose $\theta=0$, from now on. Also, scaling $x=\eps^2\tilde{x}$, $R=\eps^{-2}\tilde{R}$, we arrive at the same system \eqref{little_rho_system} with $\eps=1$. We will therefore from now on assume $\eps=1$. Altogether, we shall study 
\begin{equation}
\rho'''- \rho'(1-R(\rho,-\rho'))=0, \qquad \frac{d}{dx}\begin{pmatrix} \rho \\ \rho_1 \\ \rho_2 \end{pmatrix} =  \begin{pmatrix} 0 & 1 & 0\\ 0 & 0 & 1\\ 0 & 0 & 0 \end{pmatrix} \begin{pmatrix}\rho \\ \rho_1 \\ \rho_2  \end{pmatrix} + \begin{pmatrix} 0 \\0\\ \rho'(1-R(\rho,-\rho'))\end{pmatrix}.
\label{simp}
\end{equation}

We now collect some basic properties of \eqref{simp}. First, the reflection symmetry in the original equation,  $u \mapsto v$ and $x \mapsto -x$, induces a reversibility symmetry in  \eqref{simp}. Define the reverser acting on the phase space $M:\R^3\to\R^3$, $(\rho,\rho_1,\rho_2)\to (\rho,-\rho_1,\rho_2)$. Then, if $\underline{\rho}(x)=(\rho(x),\rho_1(x),\rho_2(x))\in\R^3$ is a solution to \eqref{simp}, then $M\underline{\rho}(-x)\in\R^3$ is a solution as well. A particular role is played by the fixed point set  $\text{Fix}\ M := \{( \underline{\rho} \ | \ \rho_1 = 0\}$.  Whenever a trajectory intersects $\text{Fix}\ M$, its reflection provides an extension of the trajectory. Also, $\text{Fix}\ M := \{( \underline{\rho} \ | \ \rho_1 = 0\}$ divides the phase space into two half spaces, $\rho_1>0$ and $\rho_1<0$. The flow restricted to each of these two half spaces possesses a Lyapunov function $V(\underline{\rho})=\rho$, which is strictly increasing in $\rho_1>0$ and strictly decreasing in $\rho_1<0$.

\begin{Lemma}\label{lemma:symmetry}  A non-equilibrium solution $\underline{\rho}(x) = (\rho(x),\rho_1(x),\rho_2(x))^T$ to \eqref{simp} cannot have more than two intersections with $\text{Fix}\ M$.  Moreover, $\underline{\rho}$ is 
\begin{enumerate}
 \item periodic if and only if it has two intersections with $\text{Fix}\ M$;
 \item homoclinic if and only if it is bounded and has one intersection with $\text{Fix}\ M$;
 \item heteroclinic if and only if it is bounded and does not have any intersections with $\text{Fix}\ M$.
\end{enumerate}
\end{Lemma}
\begin{Proof}
Assertions (i) and (ii) follow from reversibility and  from the fact that the system possesses a Lyapunov function in the complement of $\text{Fix}\ M$. The 'if' part is obtained readily by reflecting the solution using $M$ and continuing at intersection points. The 'only if' part follows from the fact that homoclinic and periodic solutions are recurrent, hence cannot be contained in the complement of $\text{Fix}\ M$, where the system has a strict Lyapunov function. Two intersections immediately imply periodicity, such that more than two intersections are not possible. One intersection together with boundedness and monotonicity of a half trajectory implies that the limit of $\rho$ exists, 
$\lim_{|x|\to\infty}\rho(x)=\rho_\infty$. Therefore, the $\omega$-limit set of the trajectory is contained in the plane $\rho=\rho_\infty$. Invariance of the $\omega$-limit set implies that $\rho_1=0$ on the $\omega$-limit set, and, subsequently, $\rho_2=0$, which proves the claim.
\end{Proof}
We will rely on this lemma primarily in Section \ref{section:asymmetric} when constructing homoclinic and heteroclinic orbits using qualitative methods.

\subsection{Aggregate sensing}\label{section:simplified}
In this section, we consider the turning rate  $r(u,v) = g(u+v)$, $g(w)=(w-1)(\gamma w-1)$. Much of the (elementary) analysis here is similar to \cite{fuhrmann} where the system \eqref{myxo_pde} with nonlinearity \eqref{e:rs} was considered with a different modeling background. 

\begin{figure}[h] 
\includegraphics[scale=0.27]{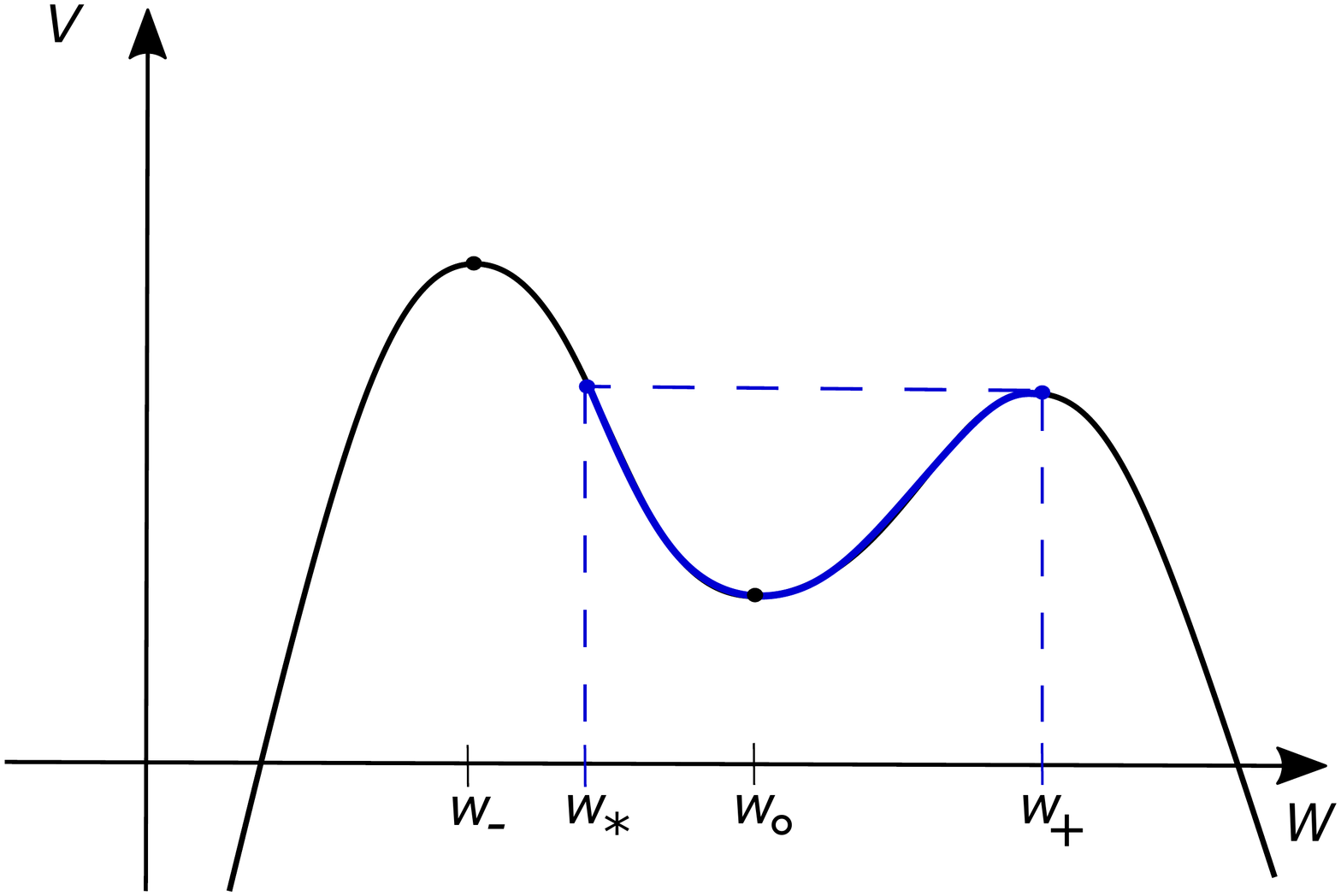}\hfill\includegraphics[scale=0.27]{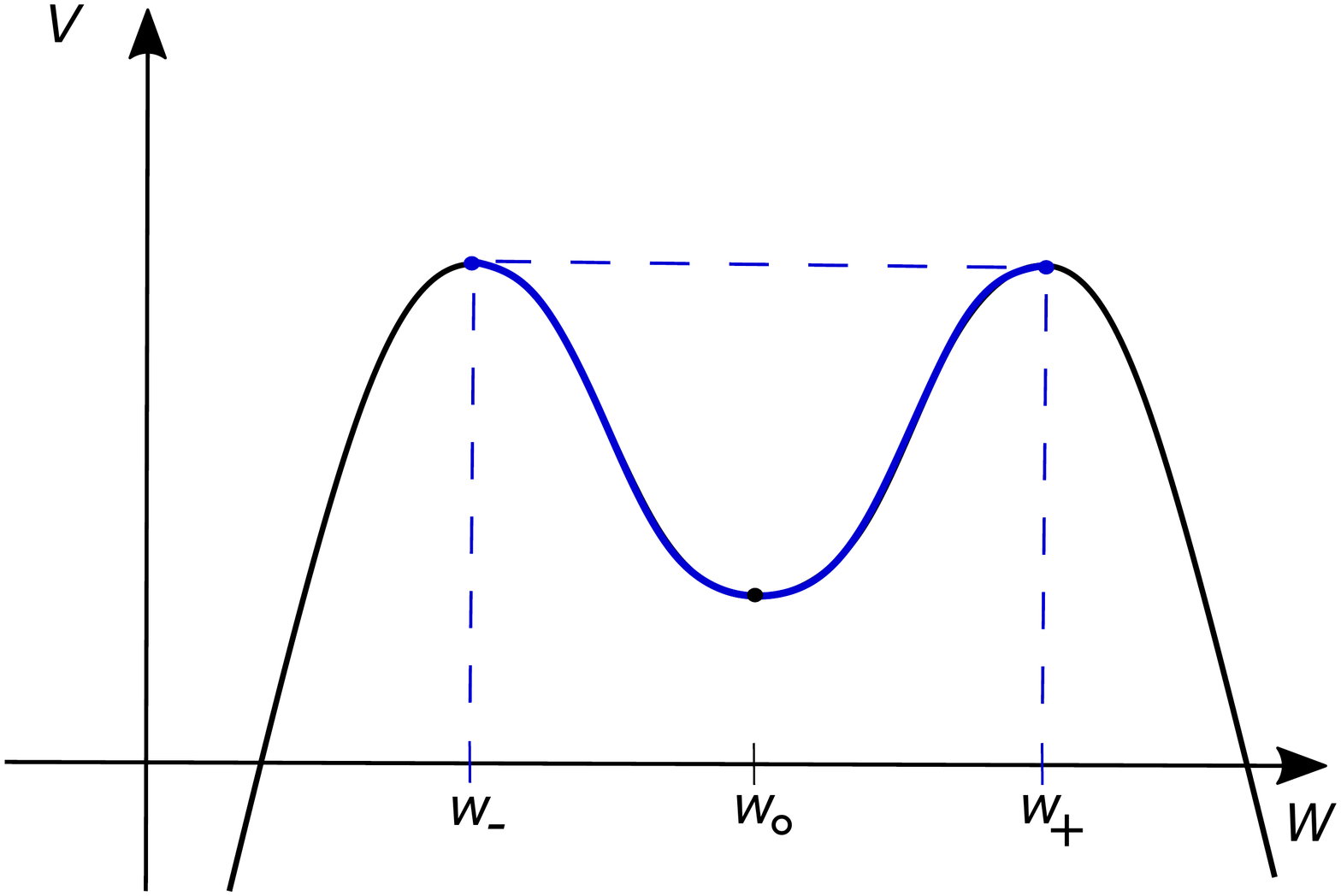}
\caption{The Hamiltonian potential, $V(w)$, with region spanned by homoclinic (left) and heteroclinic (right) orbit.}
\label{hamiltonianhomo}
\end{figure}

\begin{figure}[H]
\centering
  \begin{subfigure}[b]{0.31\textwidth}
    \centering
%   $W(\rho,c)$ for $\rho_{\pm}^-$
    \includegraphics[width=\textwidth]{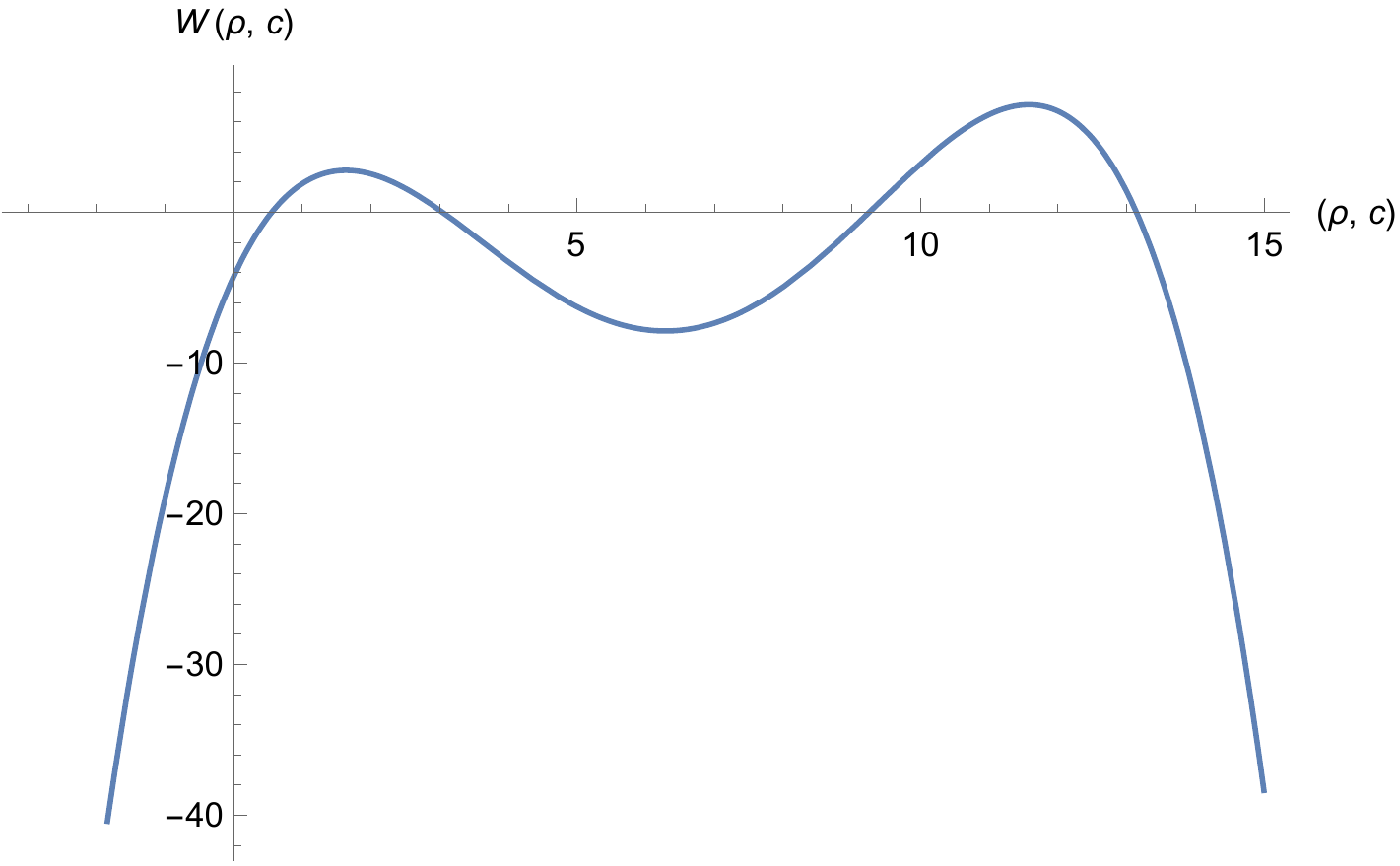}
    \caption{$\gamma = 1/8, c= 1.31$}
    
  \end{subfigure}\hfill
    \begin{subfigure}[b]{0.31\textwidth}
   \centering
%   $W(\rho,c)$ for $\rho_{\pm}^+$
    \includegraphics[width=\textwidth]{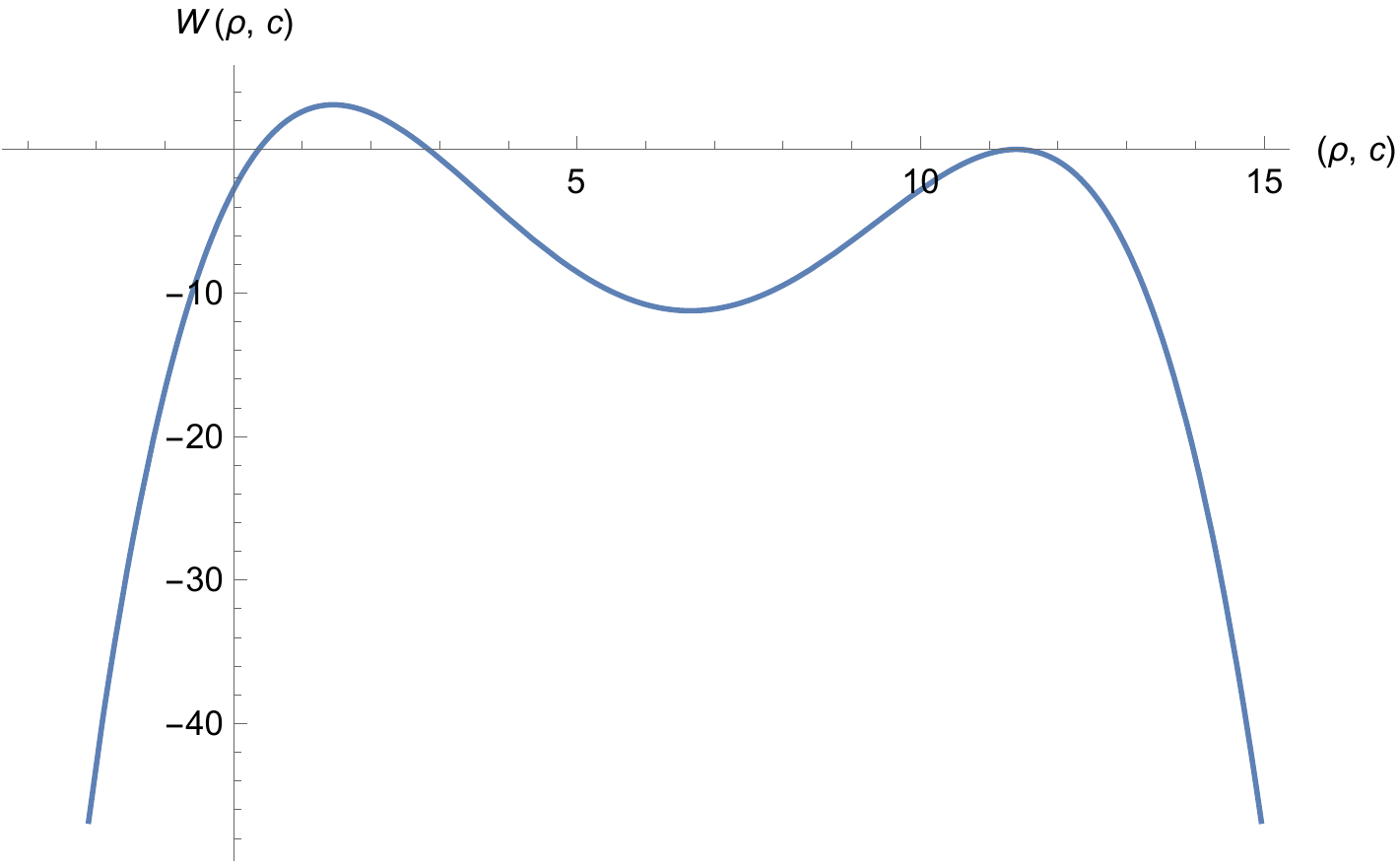}
    \caption{$\gamma = 1/8, c=2$}
   
  \end{subfigure}\hfill
 \begin{subfigure}[b]{0.31\textwidth}
    \centering
%   $W(\rho,c^*)$ for Heteroclinics $\rho_{\pm}^*$
    \includegraphics[width=\textwidth]{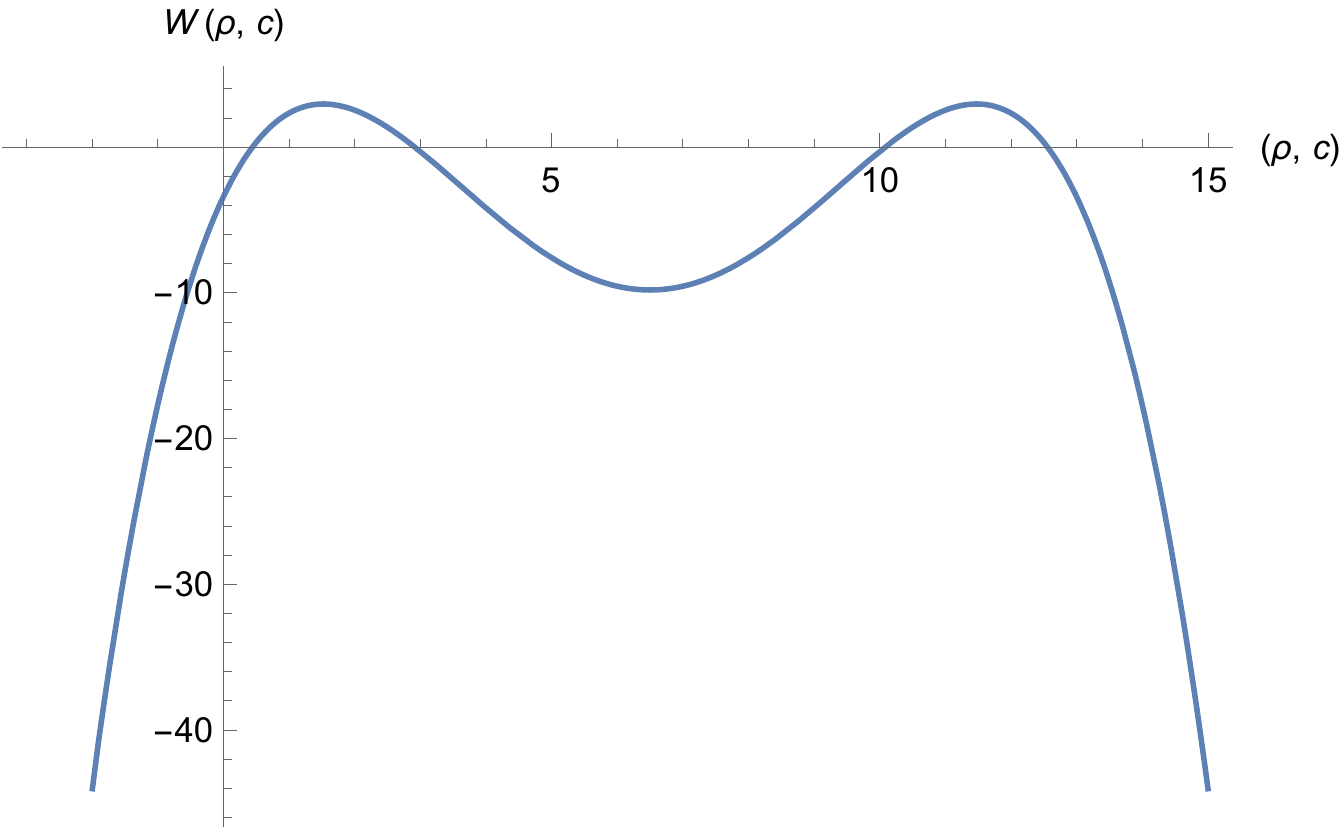}
    \caption{$\gamma = 1/8, c^*=27/16$}
    
  \end{subfigure}\\[0.2in]
  \begin{subfigure}[b]{0.31\textwidth}
    \centering
  Phase Portrait for $\rho_{\pm}^-$
    \includegraphics[width=\textwidth]{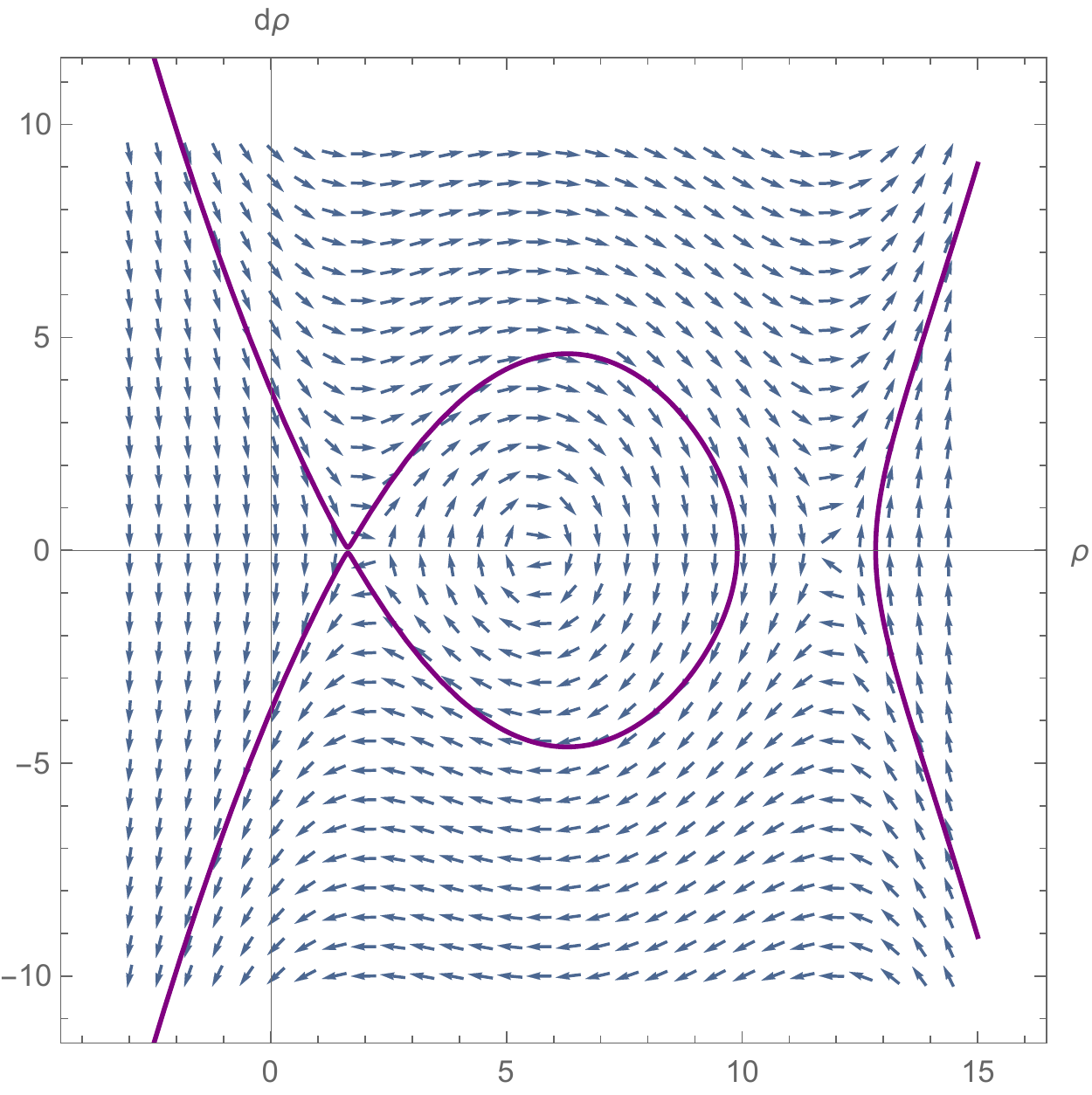}
    \caption{$\gamma = 1/8, c=1.31$}
    
  \end{subfigure}\hfill
    \begin{subfigure}[b]{0.31\textwidth}
   \centering
  Phase Portrait for $\rho_{\pm}^+$
    \includegraphics[width=\textwidth]{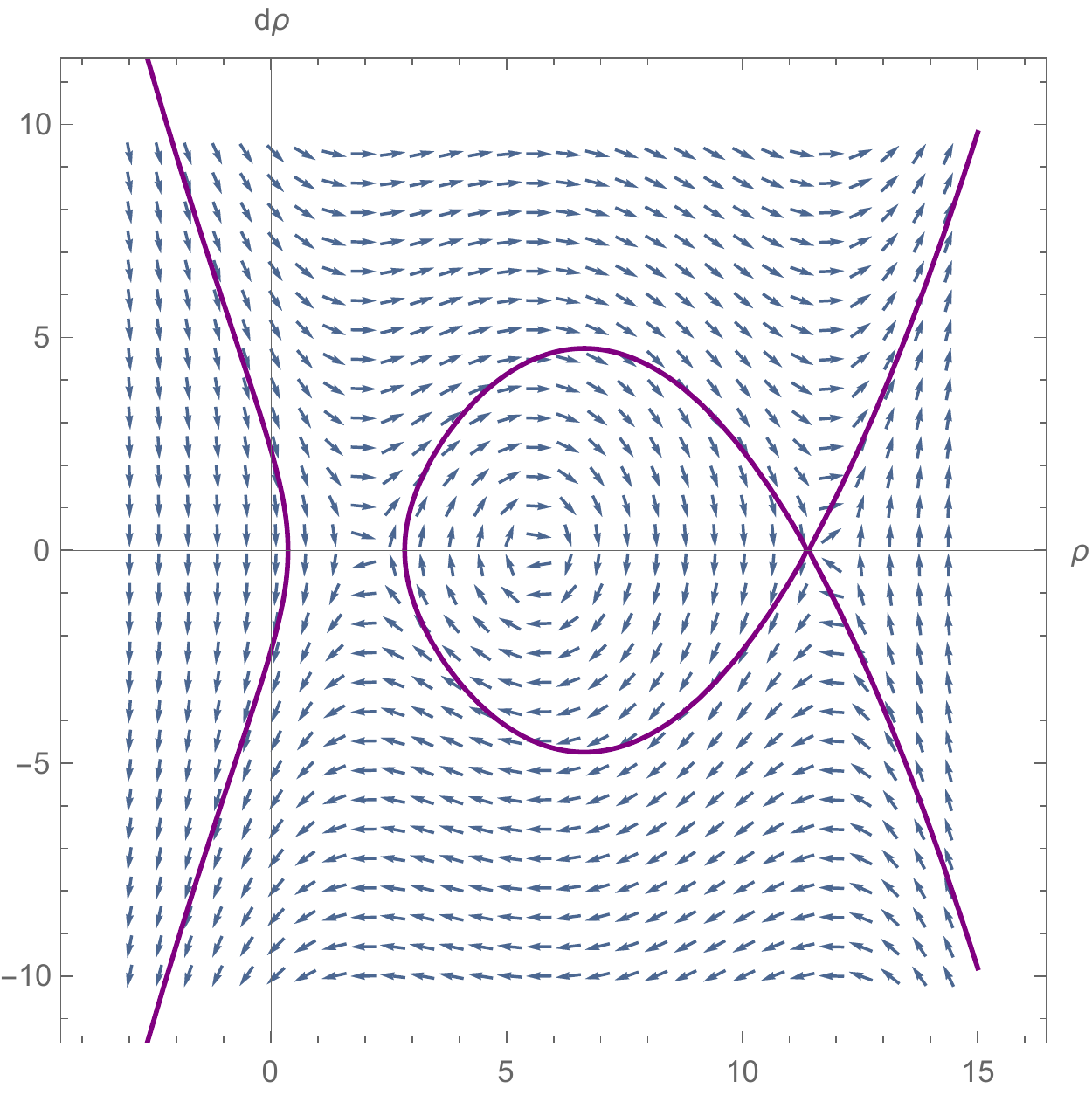}
    \caption{$\gamma = 1/8, c=2$}
   
  \end{subfigure}\hfill
     \begin{subfigure}[b]{0.31\textwidth}
   \centering
  Phase Portrait for $\rho_{\pm}^*$
    \includegraphics[width=\textwidth]{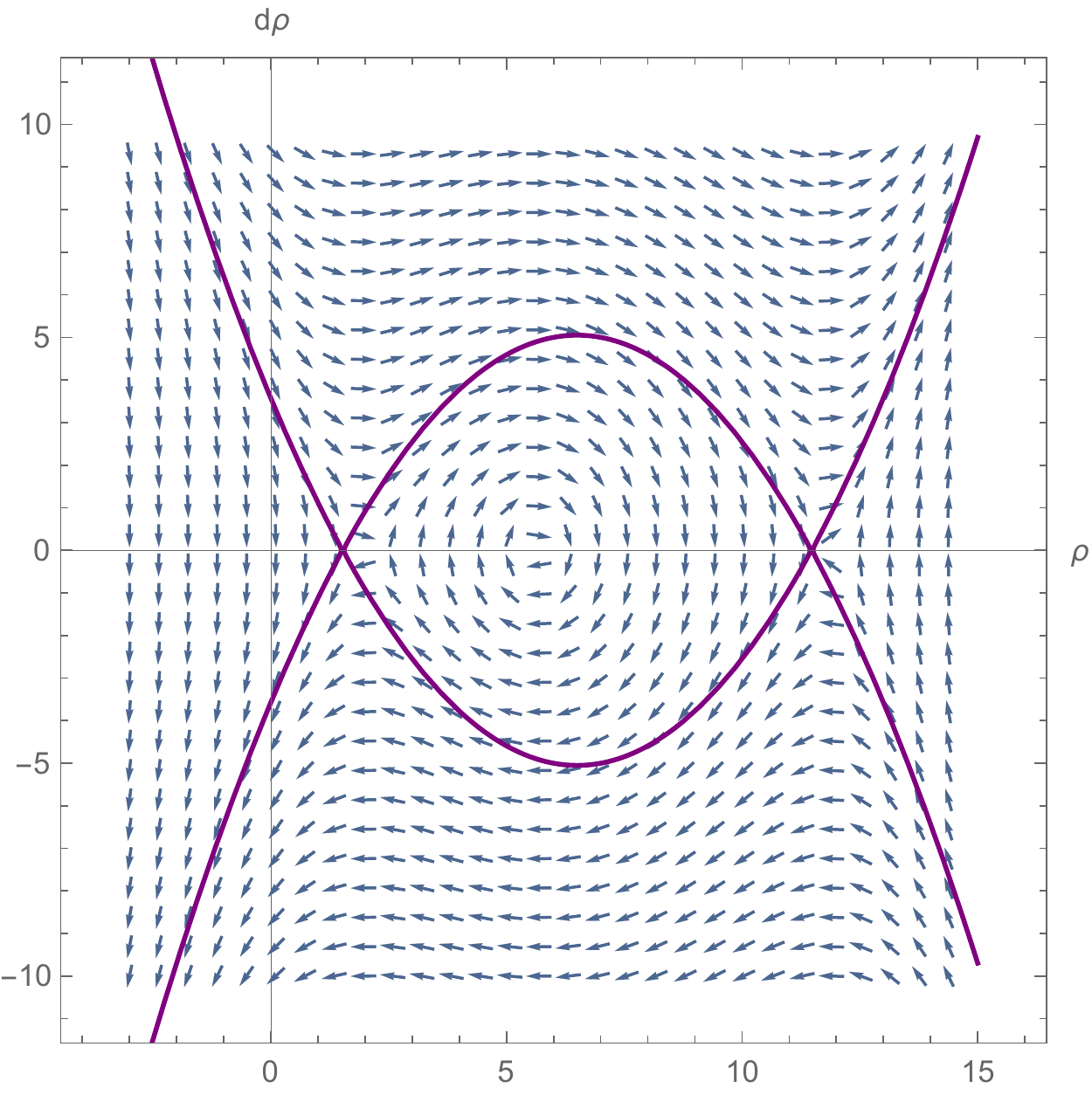}
    \caption{$\gamma = 1/8, c^*=27/16$}
   
  \end{subfigure}
\caption{Top row, left to right: Potentials $W(\rho,c)$ for cases with homoclinics to $\rho_-$, $\rho_+$, and heteroclinics, respectively.  Bottom row: Associated phase portraits.}
\label{homophase}
\end{figure}

\begin{figure}[H] \centering{
\raisebox{-1.2in}{\includegraphics[width=0.55\textwidth]{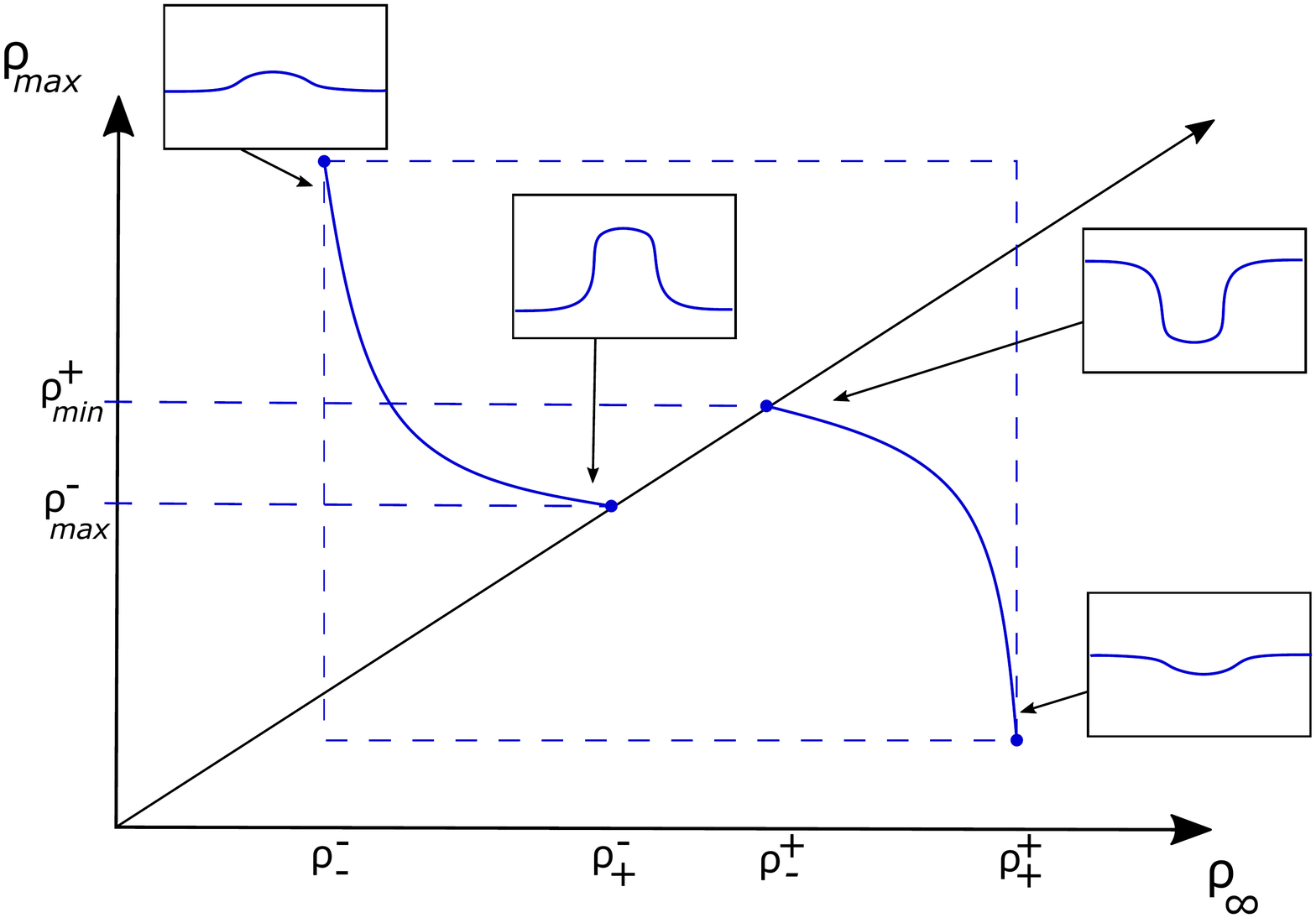}}\qquad\qquad%
\begin{minipage}{0.25\textwidth}
\includegraphics[width=\textwidth]{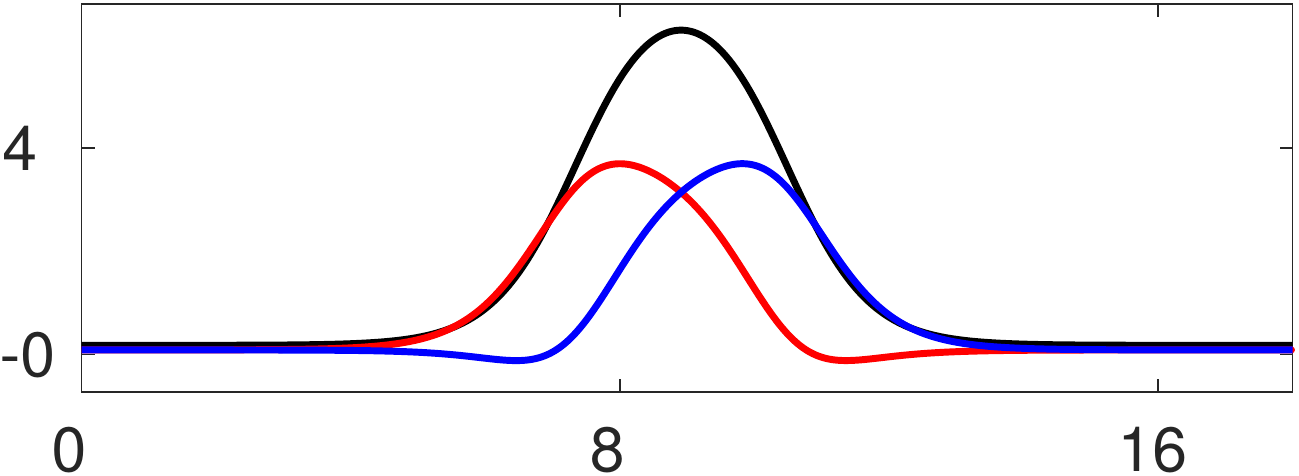}\\
 \includegraphics[width=\textwidth]{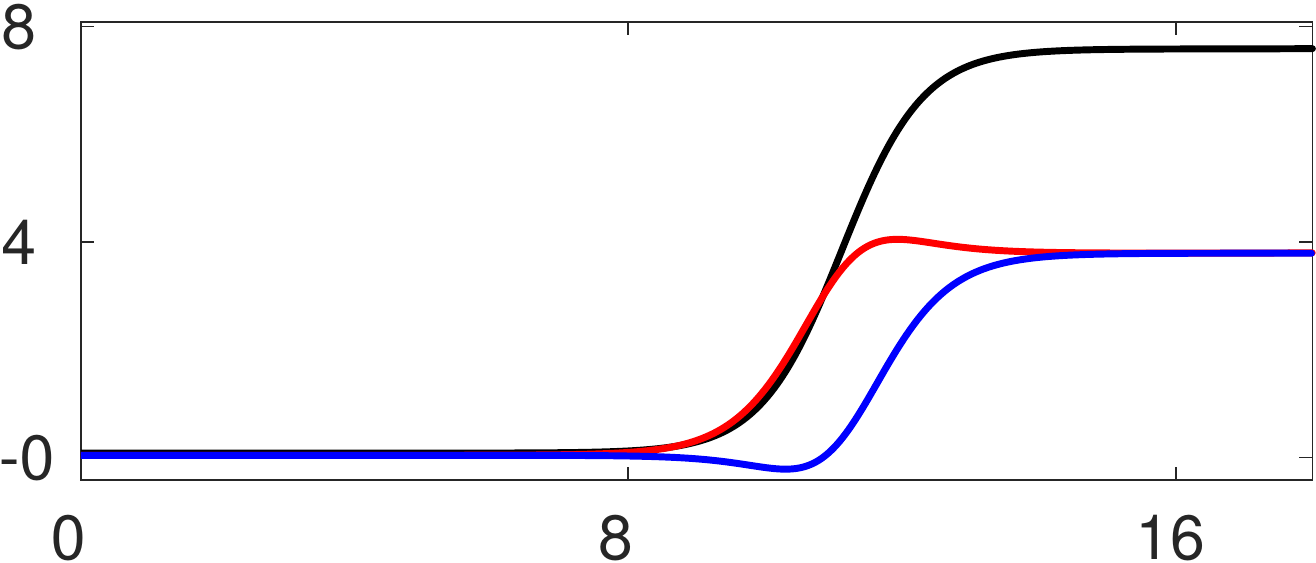}\\
 \includegraphics[width=\textwidth]{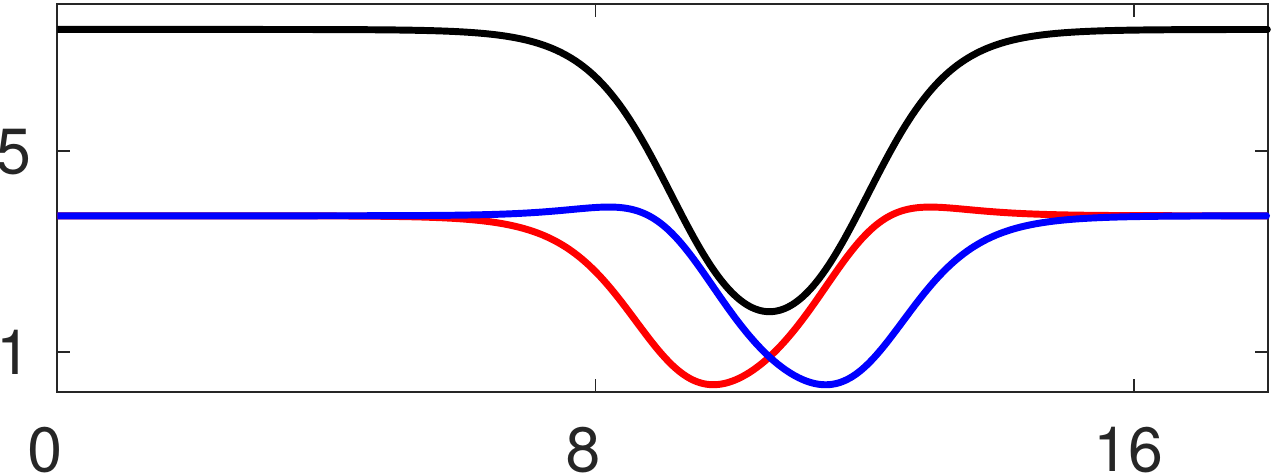}\\
\end{minipage}}
\caption{Maxima and minima of homoclinic orbits as a function of the background state (left). Plots of density profiles $u$ (blue, left-traveling), $v$ (red, right-traveling), and $\rho=u+v$ (black) for $\gamma=0.15$, $c=-0.5,-0.234,0.2$ from top to bottom (right). Note that concentrations of inward traveling populations peak at the boundary of high-density regions.}
\label{rhomax}
\end{figure}

The new turning rate, as a function of $u+v$, is described by a simplified quadratic $g(u+v) = ((u+v)-1)(\gamma(u+v)-1)$. Now we have a modified turning rate in terms of $\rho$, $R(\rho) = -2g(\rho)$, which we incorporate into equation (\ref{rho_system}). This allows us to rewrite equation (\ref{little_rho_system}) as
\begin{equation}
\rho''' = \rho'(1-R(\rho)).
\label{little_new_system}
\end{equation}
We integrating both sides with respect to $x$, and apply the chain rule to the right hand side to obtain
\begin{equation}
\rho'' = \rho-\bar{R}(\rho) + c,
\label{rhopendulum}
\end{equation}
where ${d \over d \rho}\bar{R}(\rho) = R(\rho)$, and $c$ is a constant of integration. With the specific quadratic form of $g$, this equation is given explicitly through
\begin{equation}
\rho'' = {2\over3}\gamma\rho^3 - \rho^2(\gamma+1) + 3\rho+ c.
\end{equation}
As a second order nonlinear pendulum equation, this second-order equation possesses a first integral, 
\begin{equation}\label{e:1st}
H={1\over 2}(\rho')^2 -W(\rho,c),\qquad {d \over d \rho}W(\rho) =\bar{R}(\rho),\qquad  W(\rho, c) = {1\over6}\gamma\rho^4 - {1\over 3}\rho^3(\gamma+1) + {3 \over 2}\rho^2+ c\rho.
\end{equation}
One can now proceed and find solutions explicitly, solving the corresponding first-order equation by integration. We find it more convenient to discuss solutions to this equation qualitatively. Since level sets of $H$ are invariant, orbits on bounded non-critical level sets are periodic, and orbits on bounded parts of critical level sets are heteroclinic or homoclinic. For convenience, we state the main geometric property of $W$ that we will rely on as a lemma. 
\begin{Lemma}
Consider  $w'' + V'(w) = 0$and suppose that $V$ has 3 non-degenerate critical points $w_- < w_0 < w_+$,  $V''(w_{\pm}) < 0$ $V''(w_0)>0$, $V(w_-)>V(w_+)$; see Figure \ref{hamiltonianhomo}. 
Then the set of bounded solutions consists of the three equilibria, a family of periodic orbits limiting on the equilibrium $w_0$ and a homoclinic orbit to $w_+$. When $V(w_-)=V(w_+)$, the set of bounded solutions consists of the three equilibria, a family of periodic orbits limiting on $w_0$ and a heteroclinic loop between $w_-$ and $w_+$. 
Moreover, the homoclinic orbits possess a unique maximum $w_*$ determined through the condition $V(w_*)=V(w_-)$. The case $V(w_-)<V(w_+)$ is analogous, interchanging $w_+$ and $w_-$ in the statement. 
\label{hamiltonianhomo1}
\end{Lemma}
The proof is an elementary analysis of level sets of $H$ and properties of the one-dimensional ODEs on those level sets.  
Applying this result to \eqref{rhopendulum} yields the following result on clusters and gaps in the case of aggregate sensing. 
\begin{Proposition}\label{p:agg}
For  $0< \gamma <2-\sqrt{3}$, there exists an interval of values $c$ such that there are three critical points $W(\rho,c)$,  $\rho_-<\rho_0<\rho_+$, satisfying the conditions of Lemma \ref{hamiltonianhomo1}, and therefore associated  homoclinic or heteroclinic solutions to $\rho_-$ and/or $\rho_+$. For each fixed value of $\gamma$, homoclinic or heteroclinic orbits exist for $c\in (c_-,c_+)$. There are two families of homoclinic orbits for $c\in (c_-,c_0)$ and $c\in (c_0,c_+)$, respectively, asymptotic to $\rho_\pm(c)$, respectively, limiting on $\rho_\pm$ at $c=c_\pm$ and on a heteroclinic loop at $c=c_0$, with $c_0=(1-6\gamma-5\gamma^2+\gamma^3)/(6\gamma^2)$.
\end{Proposition}
\begin{Proof}
Elementary algebra shows that the cubic function $\rho-\bar{R}(\rho)$ possesses three two critical points precisely when $\gamma <2+\sqrt{3}$. Choosing $-c$ between the values of these extrema gives three non-trivial equilibria. Elementary properties of the cubic and Lemma \ref{hamiltonianhomo1} now imply the result.
\end{Proof}
Summarizing those properties graphically, we define maxima and minima of homoclinic orbits asymptotic to $\rho_-(c)$ and $\rho_+(c)$, respectively, as $\rho_\mathrm{max}^-$ and $\rho_\mathrm{min}^+$, respectively.  Figure \ref{homophase} contains sample plots of potentials $W$ and phase portraits that illustrate how  Lemma \ref{hamiltonianhomo1} implies Proposition \ref{p:agg}. Figure \ref{rhomax} contains plots of amplitudes of homoclinic and heteroclinic orbits, as well as sample profiles of $u$ and $v$, exhibiting the formation of barriers, that is, regions of high concentration of inward traveling populations near the cluster boundaries. Those barriers become more localized and larger in amplitude as the parameter $\gamma$ decreases. We will revisit these effects in the case of head-on sensing. We note here, that the non-monotone structure of the solutions $u$ and $v$ can be directly inferred from the phase plots in Figure  \ref{homophase} after recognizing that  $u=(\rho-m)/2$, that is, monotonicity of $u$ is equivalent of monotonicity of the projection of the homoclinic onto the diagonal. 

%%% All figures here from Hom_Het_Phase.nb in 
%%% Simplified Turning Rate<Bacteria<REU2017 Dropbox

\begin{Remark}[Positivity]
 The solutions constructed here are not all positive. On the other hand, the simple translation $u\mapsto u+\delta u$, $v\mapsto v+\delta v$ with $\delta u,\delta v$ sufficiently large, preserves the form of the equation and ensures that all stationary solutions considered here are positive. On the other hand, however, all aggregate turning rates considered here are incompatible with the requirement that the temporal dynamics preserve positivity. Indeed, nullclines of $r(u,v)-r(v,u)$ are the diagonal $u=v$ and lines $u+v=\rho$, constant. Then, setting $v=0$, we see that the sign of $u_t$ changes (ignoring the advection term) when crossing $u=\rho$, implying that the positive quadrant is not forward invariant for the ODE. This in turn could be remedied by modifying the turning rate near $u=0$ and near $v=0$, preserving symmetry, such that $u_t>0$ for $u=u_x=0$. Performing this modification outside of the region of values $(u,v)$ assumed by homoclinics and heteroclinics would give explicit examples of positive clusters in a positivity preserving system. Since one may well argue with the physical relevance of the resulting nonlinearity, we do not carry out such a construction in detail, here, but rather think of these aggregate turning rates as an instructive simple example elucidating the general behavior in our run-and-tumble dynamics. 
 \end{Remark}
% \label{simplehomtheorem}
%Figure Inkscape homo.pdf

% edge.pdf
% cluster.pdf
% gap.pdf_tex
% 
% 
% Don't understand figure 2.4. what is plotted in the insets?
% \begin{figure}[H] \centering{
% \includegraphics[scale=0.38]{bifurcation.pdf}}
% \caption{Bifurcation diagram in $(u,v)$; dotted region indicates branch of equilibria where family of homoclinics exist, which limit on two heteroclinics.}
% \end{figure} 
% 

\begin{Remark}\label{r:-}
 We note that a completely analogous analysis is possible in the case of \emph{differential sensing}, when $g=g(u-v)$, In this case, we find $R=R(m)$ and \eqref{little_rho_system} is a second-order, integrable equation for $\rho_1=\rho'$, only. One looks for adjusting $\theta$ such that $w=0$ is an equilibrium and finds homoclinic orbits, which in turn yield heteroclinic orbits for $\rho$. 
\end{Remark}

\subsection{Head-on sensing}\label{section:asymmetric}

In the present section, we pursue turning rates from head-on sensing, \eqref{e:r}, with $p=q=2$; see \cite[Fig. 13]{scheel2016wavenumber} for a comparison of various turning rates. In contrast to  Section \ref{section:simplified}, the systems cannot be reduced to an integrable equation in any obvious fashion, nor were we able to identify an explicit first integral more generally for this system.  Despite this, the system still respects the symmetry and structure  outlined in Section \ref{section:symmetry} and Lemma \ref{lemma:symmetry}.  We shall exploit this structure along with estimates and perturbation arguments to find results mimicking those in Section \ref{section:simplified}.

Elementary algebra gives the explicit nonlinearity 
\begin{equation}
R(\rho,m) = -2 \left(\mu+\frac{(m^2-\rho^2 ) \left(\gamma \left(m^2-\rho
   ^2\right)+4\right)}{\left(\gamma (m-\rho )^2+4\right) \left(\gamma
   (m+\rho )^2+4\right)} \right)
%    -2\frac{r(u,v) - r(v,u)}{u-v} =-2\mu + 2\frac{4(\rho^2 - m^2) -\gamma(\rho^2-m^2)^2}{16 +8\gamma(\rho^2 + m^2)+\gamma^2(\rho^2 -m^2)^2}.
\label{bigR}
\end{equation}
in our third-order system
\begin{equation}
\rho''' = \rho'(1-R(\rho,\rho')),
\label{little_rho_system2}
\end{equation}
which can be written as a first-order ODE in three-dimensional phase space $\underline{\rho}=(\rho,\rho_1,\rho_2)^T$ in the standard fashion. The equation possesses the obvious line of equilibria $\rho\equiv \rho_\infty\in\R$, or $\underline{\rho}=(\rho_\infty,0,0)^T$. Linearizing at these equilibria,  we find the Jacobian
\begin{equation}
J = \begin{pmatrix} 0 & 1 & 0\\ 0 & 0 & 1 \\ 0 & 1-R(\rho,0) & 0 \end{pmatrix}.
\label{jacobian}
\end{equation}
Eigenvalues of $J$ are $\lambda_0=0$, associated with the family of equilibria, and $\lambda_{1/2}=\pm\sqrt{1-R(\rho,0)}$.

For $\gamma < \gamma^* := (4 + 8\mu)^{-1}$, $R(\rho,0) -1$ has precisely two positive roots $\rho_\pm$, given through
\begin{equation}
\rho_{\pm} =2\sqrt{\frac{1 - \gamma - 2  \gamma \mu \pm \sqrt{
        1 - 4 \gamma - 
         8  \gamma \mu}}{2\gamma + \gamma^2 + 
      2 \gamma^2 \mu}},
\label{roots}
\end{equation}
where one easily verifies that $\rho_\pm$ are real. In particular,  $R(\rho,0)$ is positive for $\rho \in (\rho_-,\rho_+)$, and negative for $\rho\not\in [\rho_-,\rho_+]$, $\rho>0$. Accordingly,  $J$ has eigenvalues $0, \pm\eta\in\rmi\R$, when $\rho \in (\rho_-,\rho_+)$. For the boundary case $\rho = \rho_\pm$, $J$ is a  Jordan block with eigenvalue 0. Our next result establishes existence of homoclinic orbits close to $\rho\pm$. 
% We shall then show the existence of homoclinic solutions for equilibria near $\rho_+$.  

\begin{Remark}
 We emphasize that the critical value of $\gamma_*$ is \emph{not} the critical value emphasized in \cite{scheel2016wavenumber}, which marks the minimal value of $\gamma$ for which all equilibria of the turning rate $r(u,v)-r(v,u)$ are symmetric, $u=v$. Equivalently, the values $\rho_\pm$ do not mark the boundary of stability of symmetric states $u=v=\rho/2$. Rather, increasing $u=v=\rho/2$ past $\rho_-/2$, or decreasing below $\rho_+/2$, creates a pointwise, stationary absolute instability in the sense that the Green's function admits a pinched double root at $\lambda=0$; see Section \ref{section:direct}.
\end{Remark}

\begin{Theorem}\label{theorem:perturbation} Suppose $0<\gamma <\gamma_*=(4+8\mu)^{-1}$. Then there exist two families of homoclinics $\rho_*^\pm(x;\rho_\infty)$,  of \eqref{little_rho_system2}, even in $x$, limiting on $\rho_\pm$, with asymptotic equilibria $\rho_\infty\in (\rho_--\delta,\rho_-)$ and $\rho_\infty\in (\rho_+,\rho_++\delta)$, respectively, for some $\delta>0$. Moreover, $\min_x\rho^*_+(x;\rho_\infty)<\rho_\infty$ and $\max_x\rho^*_-(x;\rho_\infty)>\rho_\infty$.
\end{Theorem}

\begin{Proof} We focus on the analysis near $\rho_+$; the analysis near $\rho_-$ is identical. 

We expand $R$ near  $(\rho,\rho') =(\rho_+, 0)$ using reversibility, and find a nonlinearity
\begin{equation}
\rho'(R(\rho, \rho_1)-1 = \alpha_{1,0} (\rho-\rho_+)\rho_1 + \rmO(\rho^2\rho_1).
\label{Rtaylor}
\end{equation}
The leading-order coefficient is obtained through  $\alpha_{1,0} = \partial_a (1-R(a,0)|_{a = \rho_+})>0$ such that equilibria  $(\rho,\rho_1,\rho_2)^T = (\rho_+ + \delta,0,0)^T$ are hyperbolic. We next scale, setting  $\rho(x)=\rho_++\epsilon \psi(\xi) $ for any $\epsilon > 0$, where $\xi = \sqrt{\epsilon\alpha_{1,0}}x$ is a scaled independent  variable. Substituting the scaling into \eqref{Rtaylor},  we find
\begin{equation}
\psi'''  = \psi\psi' +  \rmO(\epsilon).
\label{psis}\end{equation}
Formally setting $\epsilon = 0$ we readily find  a one-parameter family of explicit homoclinic solutions, 
\begin{equation}
\psi(\xi) = \psi_\infty(1- 3\text{sech}^2(\psi_\infty^{1/2} \xi/2)),
\label{sech}
\end{equation}
with $\lim_{|\xi|\to\infty} \psi(\xi) = \psi_\infty >0$. It remains to show that this family of homoclinic orbits persists for $\epsilon >0$, sufficiently small.

Therefore, notice that \eqref{psis} defines a smooth family of vector fields that possess a line of normally hyperbolic equilibria near $\psi_\infty=1$. The family of strong stable manifolds to this family of equilibria therefore depends smoothly on parameters. In the three-dimensional phases space, each of the strong stable manifolds, given by \eqref{sech}, intersects $\mathrm{Fix}\,M =\{\rho_1=0\}$, the fixed point space of the action of the reversibility. The intersection occurs at $\xi=0$ in \eqref{sech}. The tangent space to the strong stable manifold is spanned by the vector field at $\xi=0$, $(\psi',\psi'',\psi''')^T|_{\xi=0}$. Since $\psi''|_{\xi=0}\neq 0$, this tangent space is transverse to $\mathrm{Fix}\,M $. In other words, the intersection of strong stable manifolds and $\mathrm{Fix}\,M $ is transverse in $\R^3$ and therefore persists for small $\epsilon$.  This concludes the proof of Theorem \ref{theorem:perturbation}.
\end{Proof}

\begin{Remark}[Traveling clusters and gaps]
 One can of course perform a completely analogous analysis for traveling clusters, that is, stationary solutions in a comoving frame of speed $c\sim\pm1$. One then finds similar results on bifurcation of clusters and gaps, now near the critical mass concentration where $R(\rho,0)=0$, that is, near the points where asymmetric equilibria, $m\neq 0$,  bifurcate from the symmetric branch $m=0$. Indeed, in a comoving frame of speed $c=1+\sigma$ with $|\sigma|\ll 1$, we find, $\varepsilon=1$ in \eqref{rho_m_0}, 
 \begin{equation}
\begin{aligned}
0&= \rho' + m + \rho +\sigma \rho +\theta,\\
0 &=  m'' + m' + \rho' +mR(\rho,m)+\sigma m'.
\end{aligned}
\label{rho_m_01}
\end{equation}
 Substituting the expression for $\rho'$ from the first equation into the second equation, we obtain a three-dimensional ODE. Linearizing at a solution $m=0,\rho=\theta$, $\sigma=0$, where $R(\theta,0)=0$, we readily find an algebraically double eigenvalue at the origin in addition to a stable eigenvalue -1; see \cite{fs,ssmorse} for the relation between the wave speed $c$ as the group velocity and the bifurcation structure in spatial dynamics. Center-manifold reduction, expanding the nonlinearity, and exploiting the wave speed correction $\sigma$ as additional unfolding parameter, one then readily constructs homoclinic orbits bifurcating from the homogeneous equilibrium. A more comprehensive analysis of the existence and stability of such traveling structures, extending the work in \cite{fuhrmann} would be interesting but beyond the scope of the present work. 
\end{Remark}

As in the previous section, we refer to the family of homoclinic orbits with background states $\rho_\infty>\rho_+$ as gaps and to the family of homoclinic orbits with background states $\rho_\infty<\rho_-$ as clusters. The next natural question is how far this family of cluster (or gap) solutions can be extended. 
In order to address this global, non-perturbative question, we shall exploit properties established in  Section \ref{section:symmetry}, together with more subtle properties of the homoclinic orbits. 

\begin{Lemma}\label{l:mm}
For all $0<\gamma<\gamma_*$, there exists constants $C_\pm(\gamma)>0$ such that for any $\rho_\infty$ and any  $\rho(x)$, homoclinic orbit to the background state $\rho_\infty\geq 0$, we have the following a priori estimates,
\begin{align*}
\sup_x|\rho(x)|&\leq \max\{C_+(\gamma),\rho_\infty\},\\
\inf_x|\rho(x)|&\geq \min\{C_-(\gamma),\rho_\infty\}.
\end{align*}
In particular, extrema of $\max (u,v)$ at finite $x$ take values in a compact set, bounded away from the coordinate axes $u=0$ and $v=0$. As a consequence, we have that 
\[
 \|(u,v)\|_{BC^3}\leq C(\rho_\infty,\gamma),
\]
for some continuous function $C$.
\end{Lemma}
\begin{Proof}
 We use the fact that the equation for $\rho$ is equivalent to the equations for $u$ and $v$. First, an elementary calculation shows that the total turning rate satisfies $\frac{1}{2}m^2R=(u-v)(-r(u,v)+r(v,u))<0$ for $u\neq v$, except for a compact region $G_0\subset \{u,v>0\}$; see also \cite{scheel2016wavenumber}. A homoclinic orbit in the equation for $\rho$ corresponds to a homoclinic orbit to $u=v=\rho_\infty/2$ in 
 \begin{align*}
  u_{xx}+u_x-r(u,v)+r(v,u)&=0,\\
  v_{xx}-v_x+r(u,v)-r(v,u)&=0.
 \end{align*}
 Now consider $M=\max\{\sup_x u(x),\sup_x v(x)\}$. If the maximum is given by a supremum that is not attained, it equals $\rho_\infty/2$, and thus establishes the first inequality using $\rho=u+v$. If the maximum is given by a supremum that is attained, we may without loss of generality assume that $u(0)\geq u(x)$ and $u(0)\geq v(x)$ for all $x$. Therefore $u_x(0)=0$, $u_{xx}(0)\leq 0$, and $u(0)-v(0)\geq 0$, which implies either $u(0)=v(0)$ or $r(u,v)-r(v,u)\leq 0$, that is, $u,v\in \overline{G_0}$. In the former case, $u_x(0)=v_x(0)=0$ and $u=v$, that is, the solution is at an equilibrium at $x=0$, a contradiction. In the latter case, we established that $(u,v)$ are a priori bounded, in $G_0$. This proves the first inequality. The proof of the second inequality is analogous. The a priori bounds for $u$ and $v$ are then obtained readily exploiting the bounds on $r(u,v)$, writing for instance $u_{xx}+u_x-u=r(u,v)-r(v,u)-u$, using that the right-hand side is bounded in $BC^0$, and that the left-hand-side defines an invertible operator. Boot strapping then gives bounds on higher derivatives.
\end{Proof}
One can in fact refine the a priori bounds inspecting the extrema of $u-v=m=-\rho'=-\rho_1$, that is, inflection points of $\rho$, using that
\[
 \rho_1''=\rho_1\left(1-R(\rho,\rho_1)\right).
\]
Here, the maximum principle implies that the extrema of $\rho_1$ are contained in $\bar{G}$, where 
\begin{equation}\label{e:g}
 G=\{(\rho,\rho_1)| \rho>0,\ R(\rho,\rho_1)>1\}.
\end{equation}
Note that trivially $G\subset G_0$. 
\begin{Lemma}\label{l:g}
 We have $G\subset \{(\rho,\rho_1)|\rho>\rho_-\}$.
\end{Lemma}
\begin{Proof}
 The proof consists of straightforward but slightly tedious algebra. One finds the boundary of $G$ in $m=\rho_1>0$ given through curves
 \begin{equation}
\rho_G^\pm(m)=
   \sqrt{\frac{\gamma^2 m^2 \tilde{\mu}\pm 4
   \sqrt{-\gamma^3 \tilde{\mu}^2m^2-2 \gamma^2 m^2 \tilde{\mu}-4 \gamma \tilde{\mu}+1}+2 \gamma \left(m^2-2\tilde{\mu}\right)+4}{\gamma ( \gamma \tilde{\mu}+2)}},
 \end{equation}
where $\tilde{\mu}=2\mu+1$. Clearly, $\rho_G^+(m)>\rho_G^-(m)$, and for $m>0$, 
\[
\mathrm{sign}\,\left( \frac{\rmd}{\rmd m}\rho_G^-(m)\right)=\mathrm{sign}\, \left(\sqrt{-\gamma^3\tilde{\mu}^2 m^2-2
   \gamma^2 m^2\tilde{\mu}-4 \gamma\tilde{\mu}+1}+2\gamma\tilde{\mu} \right)
%    {\sqrt{-4 \gamma^3 m^2
%    \mu ^2-4 \gamma^2 m^2 \mu -8 \gamma \mu +1}
%    \sqrt{\frac{\gamma^2 m^2 \mu -2 \sqrt{-4 \gamma^3 m^2 \mu ^2-4
%    \gamma^2 m^2 \mu -8 \gamma \mu +1}+\gamma m^2-4 \gamma
%    \mu +2}{\gamma^2 \mu +\gamma}}}\right)
>0,
\]
when $m>0$. At $m=0$, $\rho_G^-(m)= \rho_-$, proving the claim.
   \end{Proof}

\begin{figure}[H]
\def\svgwidth{\columnwidth}
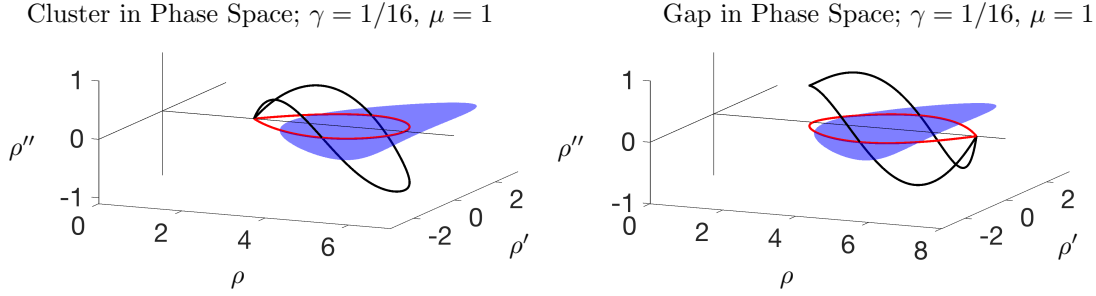
\caption{Homoclinic trajectories in phase space (black), their projections onto the $\rho$-$\rho'$ plane (red) and the region $G$ in the $\rho$-$\rho'$ plane (blue).  See Section \ref{section:hom_cont} for details on how these solutions were found numerically.}
\label{fig:phase}
\end{figure}

We present a cluster and gap solution as a trajectory in phase space together with the region $G$ in Figure \ref{fig:phase} for $\gamma = 1/16$ and $\mu = 1$. A priori bounds will allow us to continue homoclinic solutions globally, beyond the small amplitude regime in Theorem \ref{theorem:perturbation}. The following result characterizes possible end points to such a continuation.

\begin{Corollary} Let $S$ be the set of all equilibria in $[0,\rho_-)$ with homoclinic orbits and suppose there is $\rho_\infty^* \in (\partial S)\setminus \{\rho_-\}$. Then there exist two positive heteroclinic solutions $\rho^*(x)$ and $\rho^*(-x)$ connecting equilibria $\rho^*_\infty$ to $\rho_+^* \in [\rho_+,\infty)$.  
\end{Corollary}

\begin{Proof}  Consider a sequence of homoclinic orbits $\rho^n(x)$, even,  asymptotic to $\rho_\infty^n\to \rho_\infty^*$. By normal hyperbolicity of $\rho_\infty^*\in ([0,\rho_-)$, we can identify the homoclinic solutions with unstable manifolds, limiting on the one-dimensional unstable manifold of $\rho_\infty^*$. The trajectory $\rho^*(x)$ in this unstable manifold is bounded as the limit of uniformly bounded functions, not homoclinic, since not in $S$, and not periodic. It is therefore heteroclinic, by Lemma \ref{lemma:symmetry}. It remains to show that $\rho_+^*\geq \rho_+$. Since $\rho^*(x)$ is monotone and convergent, its derivative converges to zero and therefore possesses an extremum. By Lemma \ref{l:g}, this extremum lies in $G$, with a value of $\rho^*(x_0)\geq \rho_-$. This together with monotonicity of $\rho^*(x)$ establishes that $\rho_+^*> \rho_-$. It remains to show that $\rho_+^*\not\in (\rho_-.\rho_+)$. For this, notice that the linearization at equilibria in this interval is elliptic, with eigenvalues  $0,\pm\omega \rmi$. Following the proof of Theorem \ref{theorem:perturbation}, one sees that the local dynamics near the equilibrium $\rho=\rho_\infty\in (\rho_-.\rho_+)$ are
\[
\rho'''+\alpha^2\rho'+\rmO\left((\rho-\rho_\infty)^2\rho'\right)=0,                                                                                                                                                                                                                                                                                                                                                                                                                                                                                                                                                                                                                                                                                                                                                                                                                                                                                                                                                                                                                                                                                                                                                                                                                                                                                                                                                                                                                                                                                                                                                                                                                                                                                                                                                                                                                                                                                                                                                                                                                                                                                                                                                                                                                                                                                                                                                                                                                                                  \]
for some $\alpha\neq 0$. At leading order, we find a family of periodic orbits, $\rho=\rho_\infty + A\cos(\alpha (x-x_0))$. By reversibility, this family of solutions persists for the full system. Varying $\rho_\infty,A,x_0$, we find that a neighborhood of the line of equilibria is filled with periodic orbits, such that there do not exist solutions that converge to $\rho_\infty$ for $x\to\infty$. 
% 
% 
% Since $\rho_\infty$ lies in $(-\rho_-,\rho_-)$, it is normally hyperbolic with one-dimensional unstable manifold, given by the trajectory $\rho^*(x)$, $\lim_{x\to-\infty} (\rho^*)'(x) >0$.  Then,  $\rho_\infty \in (\rho^*_-,b)$ have a cluster $\rho(x)$ with $\rho'(0) =0$.  We have that the unstable subspaces for the linearization about equilibria in $(0,\rho_-)$ vary smoothly in $\rho_\infty$ within a neighborhood of $(\rho_-^*,0,0)^T$ in phase space; we use this fact, along with
% smoothness on initial conditions, to get the following continuity result:
% For any $x_f < 0$ and $\epsilon > 0$, there exists an $x_0$ and $\delta>0$ such that for all $\rho_\infty \in B(\rho_-^*,\delta)$, $\|\rho^*(x)-\rho(x-x_0)\|_{C^2(-\infty,x_f]} < \epsilon$.  Thus, by the preceding Theorem, it follows that $\rho^*$ is also bounded.  But, $\rho^*$ is neither periodic nor homoclinic, so it must be heteroclinic. Moreover, $\rho^*(x)$ must have at least one point of inflection.  One can show that $G \subseteq (\rho_-,\infty)\times \mathbb{R}$.  Thus $\rho^*(x) > \rho_-$ for all $x$ large enough.  Since $\lim_{x\to \infty}\rho^*(x)$ cannot be a elliptic equilibrium, we have that $\lim_{x\to\infty} \rho^*(x) \geq \rho_+$.  
\end{Proof}

Loosely, this corollary states that ``homoclinics limit on heteroclinics as one varies the background state $\rho_\infty$''.  This observation forms the basis for the numerical continuations of Section \ref{section:numerics}.  This corollary also gives us a sufficient condition for the existence of a heteroclinic: suppose  there is $\rho_-^*>0$ such that there does not exist a homoclinic orbit to $\rho_-^*$; then there exists a heteroclinic orbit for some $\rho_\infty\geq\rho_-^*$. 

In the following, we show how to derive existence of such heteroclinic orbits for $\gamma$ moderately small. To do this, it suffices to show that the largest interval $(\rho_-^*,\rho_-)$ for which all $\rho_\infty \in (\rho_-^*,\rho_-)$ is bounded away from $-\rho_-$ (and ideally nonnegative  since $\rho$ represents a concentration).

First, while this system has no obvious first integral, we can derive estimates through comparison with an integrable system.  Therefore, consider any function $Q(\rho)$ such that $Q(\rho) \geq R(\rho,\rho_1)$ for all $\rho$, $\rho_1$.  Then, if $\rho(x) \to \rho_\infty$ as $x \to -\infty$ and $\rho$ is strictly increasing on $x\leq x_f$, we have $\rho'' \geq \int_{\rho_\infty}^{\rho} (1-Q(a))da$ for all $x \in (-\infty,x_f)$. Multiplying by $\rho'$ and integrating in $\rho$ once more, we obtain an explicit bound on $\rho'$,
\begin{equation}
\rho' \geq \sqrt{2{\int_{\rho_\infty}^{\rho}\int_{\rho_\infty}^{b} (1-Q(a))\ da\ db}} =: q(\rho).
\end{equation}
Thus, if $G \subseteq \{(a,b)^T \ | \ |b| \leq q(\rho), a >\rho_\infty\}=:B$, then $(\rho,\rho')^T$ never enters $G$ and so $\rho''' >0$ for all $x < x_f$.  In such a case, $\rho''(x_f)$ is never zero.  This implies $\rho''(x) >0$ for all $x$ where $\rho$ is defined, and likewise for $\rho'$.  Thus $\rho(x) \to \infty$ in finite or infinite $x$.

To obtain  nonnegative solutions, we set $\rho_\infty = 0$.  The simplest nontrivial candidate for an upper bound $Q$ is 
\begin{equation}
Q(\rho) = -2\mu + \rho^2/(2+\gamma \rho^2).
\label{Q}
\end{equation}
One can use a more sophisticated bound, taking for instance the critical points of $R$ for fixed values of $\rho$ and finding local maxima. Trying this approach, we found little improvement over the estimates obtained here. In our case, one can calculate the regions $G$ and $B$ explicitly.
%The full computation of $q(\rho)$ is
% \begin{equation}
% q(\rho) = \frac{1}{\gamma}\sqrt{\rho ^2 (2 \gamma^2  \mu +\gamma^2 -\gamma)+2 \log \left(\frac{2}{\gamma  \rho ^2+2}\right)+2 \sqrt{2\gamma } \rho  \arctan\left(\sqrt{\frac{\gamma }{2}} \rho\right)}.
%    \label{little_q}
% \end{equation}
% Meanwhile, $\bar m(\rho) := \sup \{m \ |\ (\rho,m) \in G\}$ is given by
% \begin{equation}
% \bar m(\rho) = \sqrt{\rho^2 -\frac{4(2 \mu+\gamma+1)-4\sqrt{1-\gamma ^3 (2 \mu  \rho +\rho )^2-2 \gamma ^2 (2 \mu +1) \rho
%    ^2}}{2 \gamma^2  \mu +\gamma^2 +2\gamma}}.
% \end{equation}
We show a sample of plots of the regions $G$ and $B$ for a few values of $\gamma$ and $\mu = 1$, Our plots suggest that $B\subseteq G$ whenever $\gamma \in (\gamma_0,1/12)$, where $\gamma_0 \in (1/18, 1/17)$.  For all smaller $\gamma$, we found that $G \not \subseteq E$.  This is not surprising: when $\gamma = 0$, $G$ is unbounded, yet $B$ is bounded. In summary, the plots show that positive heteroclinic orbits exist for $\gamma\geq \gamma_0$. 

% plots made using blowup_solns.nb
\begin{figure}[H]
\centering
\includegraphics[width = .7\textwidth]{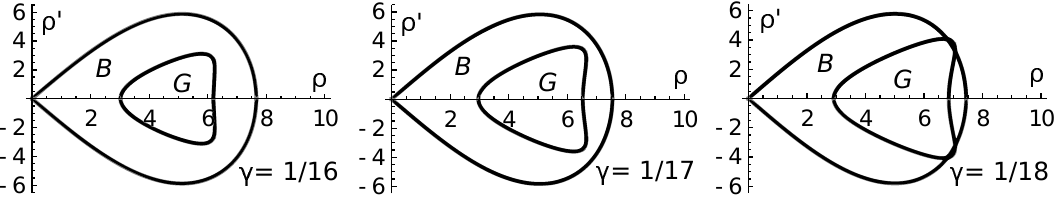}
\caption{Plot of the boundaries of $B$ and $G$ in the $\rho$-$\rho'$ plane for $\mu = 1$.}
\end{figure}
\label{goob_boog}
% We also provide a more rigorous (but less explicit or geometric) argument for the existence of heteroclinics when $\gamma <\gamma^*$ is in a neighborhood of $\gamma^*$.  Once again, we prove the existence of unbounded increasing solutions along the unstable manifolds of equilibria $\rho_\infty \in [0,\rho_-)$.
We supplement this analysis with a perturbational argument.  Take $\rho(x)$ to be in the unstable manifold of $\rho_\infty$ with $\rho' > 0$ for all sufficiently negative $x$.  This solution is unique up to translation in $x$.   As the region $G$ is empty for $\gamma \geq \gamma^*$, $\rho(x)$ is necessarily increasing and concave up for all $x$.  Thus, for such $\gamma$, $\rho$ is unbounded.  Next, take $\gamma_1 \in (0,\gamma^*)$.    Choose an equilibrium such that
$0\leq \rho_\infty < \inf_{\gamma_1 \leq \gamma \leq \gamma^*} \rho_-$.  Set $\bar a= \sup\{a \ | \ (a,b) \in G,\gamma_1 \leq \gamma \leq \gamma^*\}$, i.e. a right bound on $G$ in the plane for all $\gamma \in (\gamma_1,\gamma^*)$.  We have that $\rho(x) > 2\bar a$ for all sufficiently large $x$.  Since the flow is continuous in $\gamma$, there exists a $\gamma_2 \in (\gamma_1,\gamma^*)$ such that $(\rho,\rho',\rho'')^T$ enters a compact subset of $(\bar a,\infty)\times\mathbb{R}_+\times\mathbb{R}_+$ for all $\gamma \in (\gamma_2,\gamma^*)$.  This implies that $\rho$ has positive first, second and third derivatives for all later $x$, giving unboundedness. 

More explicit expansions in this case can be found through a bifurcation analysis, mimicking Theorem \ref{theorem:perturbation}. At $\gamma=\gamma_*=\frac{1}{12}$, the quadratic term $\rho\rho_1$ in the Taylor expansion  \eqref{Rtaylor} vanishes and is replaced by
\[
 \rho_1\left(1-R(\rho,\rho_1)\right)=\rho_1\left(\frac{27}{64}(\rho-4)^2+36\left(\gamma-\frac{1}{12}\right)+\rmO(|\rho_1^2\rho|+|\rho^3|)\right),
\]
representing, at leading order, a cubic second order equation with heteroclinic orbits when zeros are equidistant. Those heteroclinic orbits are transverse when viewed as intersections of the family of unstable manifolds of equilibria with the family of stable manifolds of equilibria, hence persist for higher-order perturbations.

\section{Numerical Computations}
\label{section:numerics}
We compute the two families of homoclinic solutions and the heteroclinic limit using an arc-length continuation method in Section \ref{section:hom_cont}.  Adapting this method, we  compute and continue heteroclinic solutions  in $\gamma$, investigating in particular the limit $\gamma \to 0$ in Section \ref{section:het_cont}.  

\subsection{Homoclinic Continuation}
\label{section:hom_cont}
\begin{figure}[h!]\begin{center}
$\gamma = 1/16$\\ 
% dx = 0.01
\includegraphics[width= 0.8\textwidth]{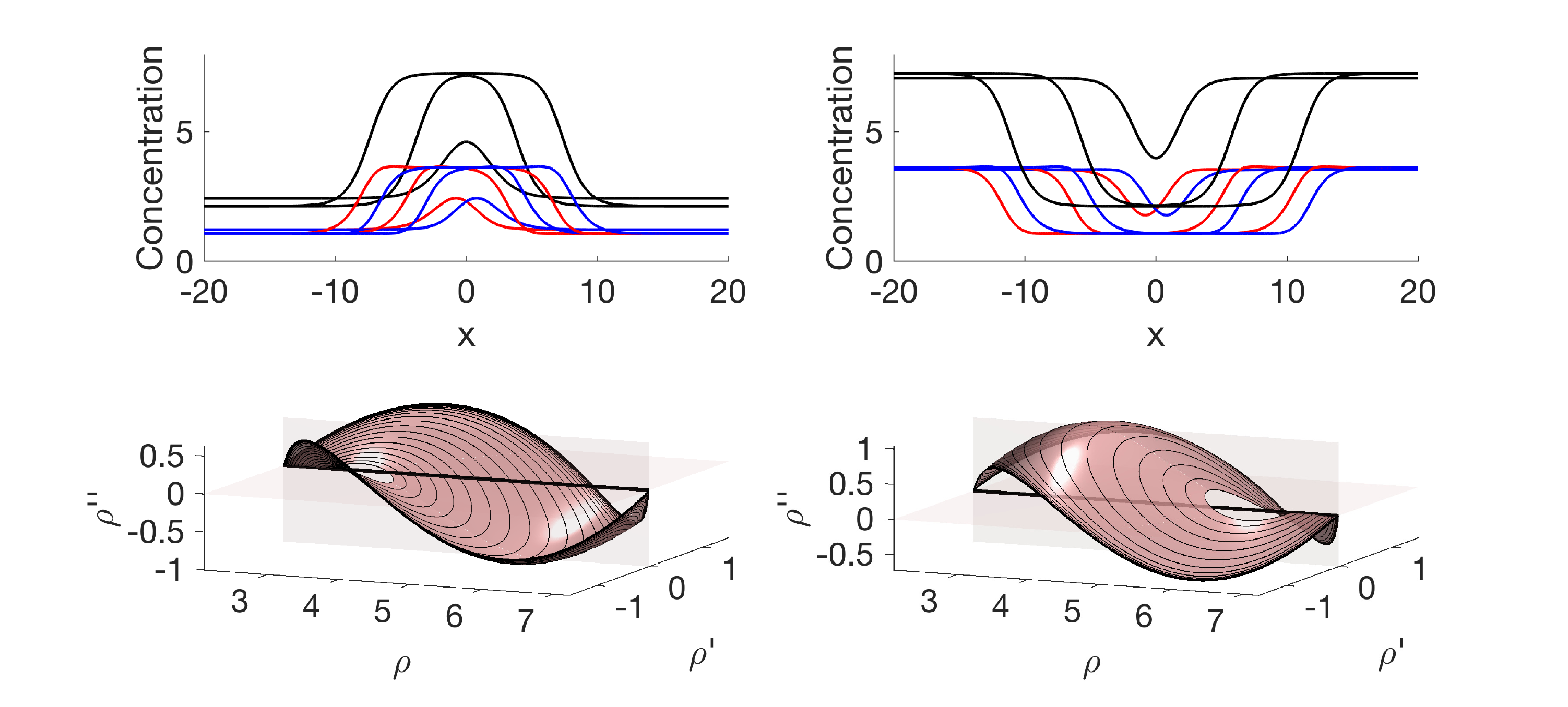}\\
$\gamma= 1/64$\\
% dx = 0.025
\includegraphics[width= 0.8\textwidth]{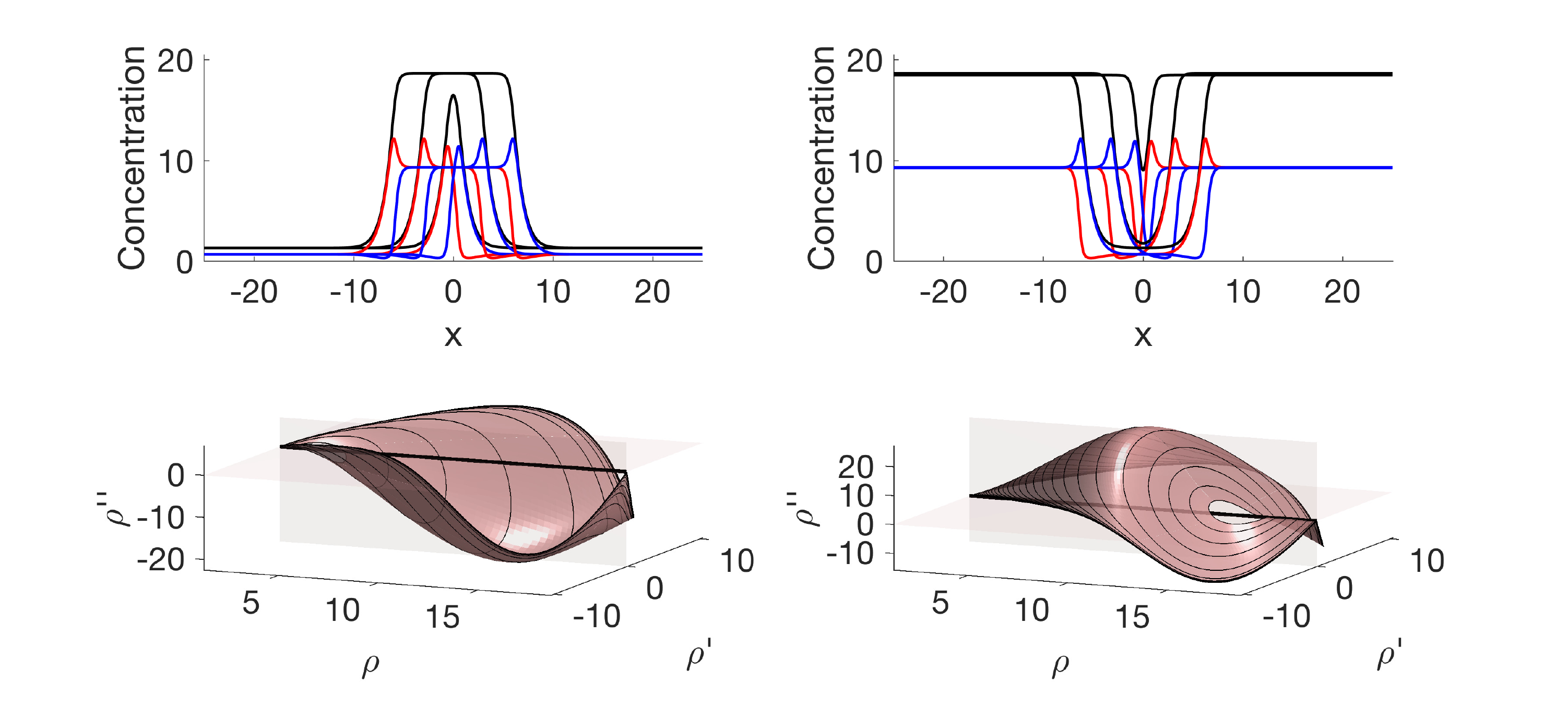}\\
$\gamma =1/256$\\
% dx = 0.01
\includegraphics[width= 0.8\textwidth]{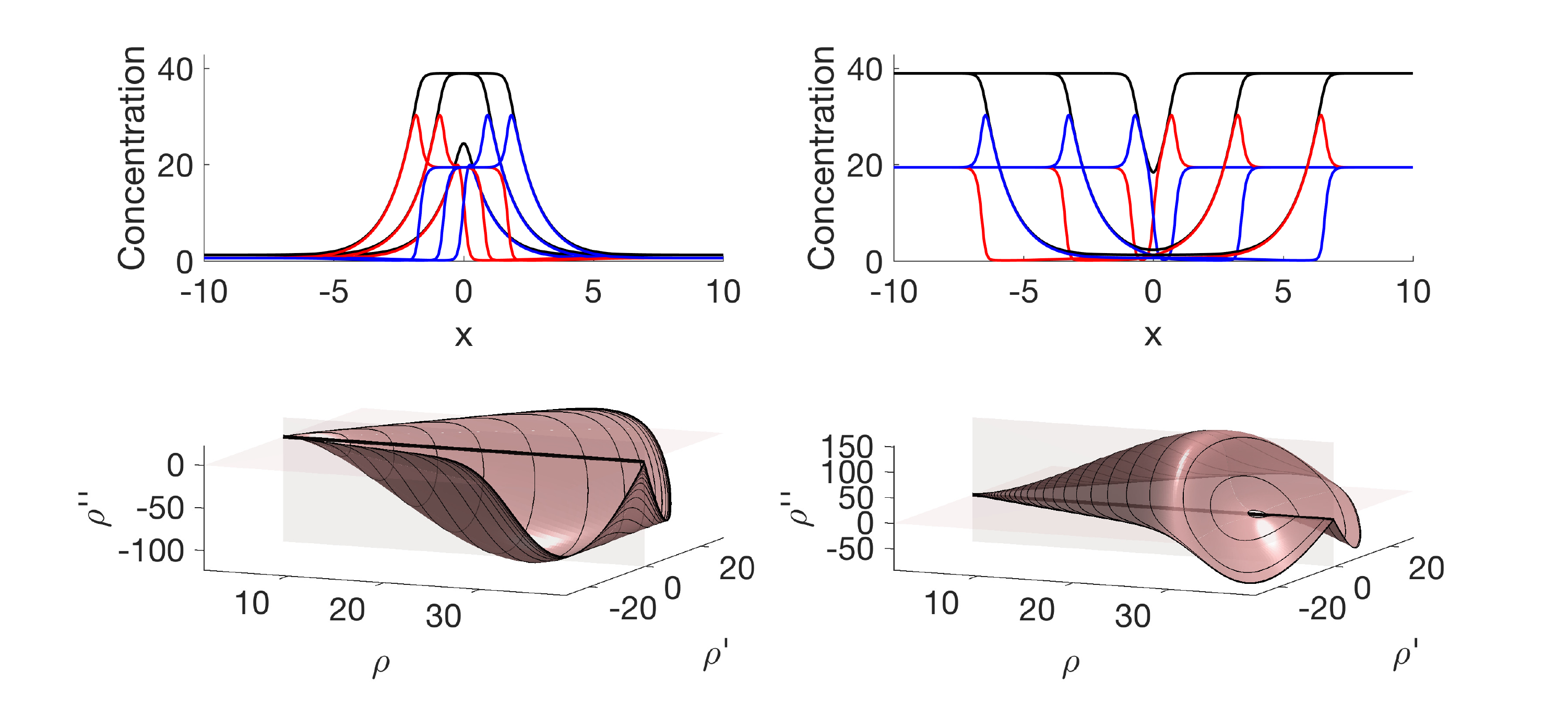}\\
\end{center}
\caption{Cluster and gap solutions, with associated phase portraits. Individual plots show the families of solutions as the background state is varied. Different plots correspond to different values of the parameter $\gamma$. Shown is the actual computational domain, grid sizes vary in $dx=0.01\ldots 0.025$. }
\label{hom_conts}
\end{figure}

For our numerical computations, we consider \eqref{little_rho_system2} with turning rate  \eqref{bigR} and set $\mu = 1$. To find homoclinic solutions, we used the arc-length continuation method to vary the background state  $\rho_\infty =\lim_{|x|\to\infty} \rho(x)$. We do this for both cluster and gap solutions.  For an initial guess, we adapt \eqref{sech} for the equation $\rho''' = \rho'(1-R(\rho,\rho'))$.  We take $\rho(x) = \rho_\pm \pm \delta(1-3\sech^2(|\alpha\delta|^{1/2} x/2))$ and $\delta> 0$ to be small, where  $\alpha := \partial_1(1-R(\rho_\infty,0))$.  We compute homoclinics on a domain $[0,L]$, using the left boundary condition $\rho'(0) = 0$. At $x=L$, we approximate the strong stable manifold through its tangent space and use the associated projection boundary conditions. Discretization uses the second-order trapezoidal rule on grid sizes $dx\sim 10^{-2}$ We present samples of families of homoclinics in Figure \ref{hom_conts} as functions of $x$, with cluster solutions on the left and gap solutions on the right.  Below these plots, one finds the corresponding foliated surface in the phase space of \eqref{little_rho_system2}: the surface is (one branch) of the union of strong unstable manifolds, thus part of the center-unstable manifold of the line of equilibria. Black lines indicate individual trajectories alias strong unstable manifolds. 

% plots made with homs_and_hets/homs/tiled_cont_plots.m, adapting code from graph_homs.m in same folder
% continuation code is homs_and_hets/homs/homoclinic_cont.m

\begin{figure}[h!]
\centering{\includegraphics[width= .85\textwidth]{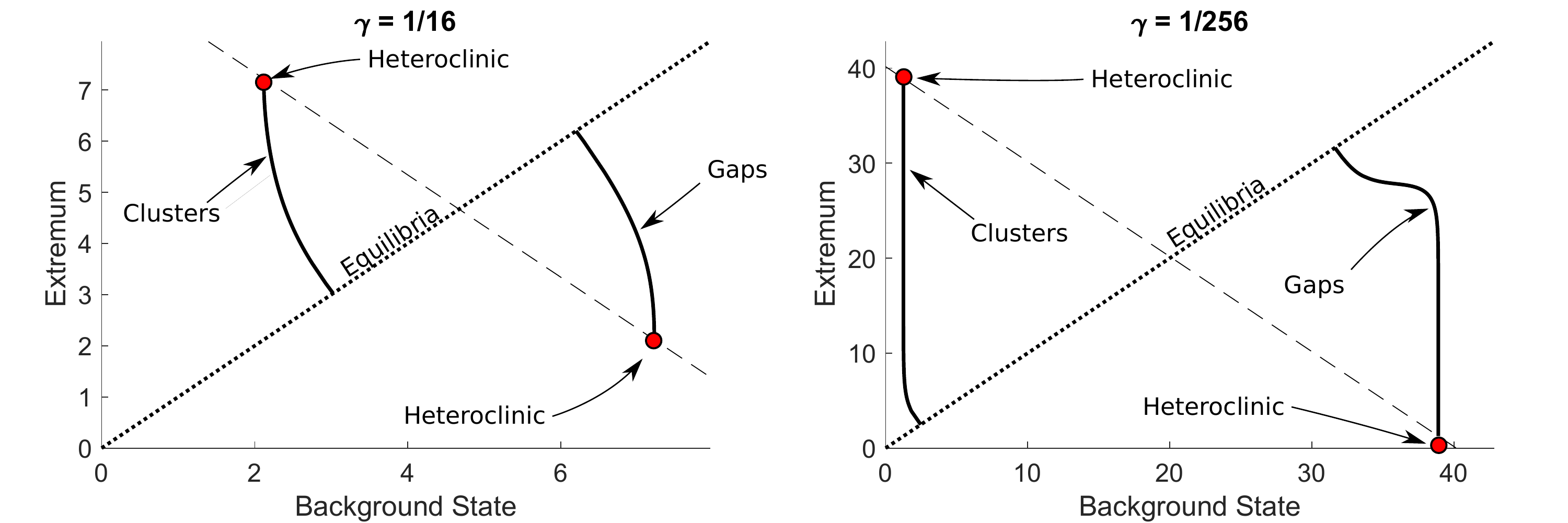}}
\caption{Maxima of clusters and minima of gaps in the continuation, plotted against the background state, for sample values of $\gamma$.}
\label{ext_v_bgst}
\end{figure}

Figure \ref{hom_conts} strongly suggests that both clusters and gaps limit on a heteroclinic orbit, that is, a cluster boundary.  Figure \ref{hom_conts} also suggests that the limiting heteroclinic is the same, up to translation and reflection in $x$ for both cluster and gap continuations.   Moreover, once the cluster or gap flattens out at $x=0$, the background state barely changes and the continuation only widens the heteroclinic plateau.   In Figure  \ref{ext_v_bgst}, we show the extreme values of these clusters as a function of their background state.  These plots serve as the numerical counterparts to Figure \ref{rhomax}.  As the cluster limits on a heteroclinic, its maximum converges to the final background state of the gap.  Similarly, the minimum of the gap converges to the background state of the cluster.  

% plots made with homs_and_hets/homs/compare_homs.m

% Subsection (1) is done in terms of code, all I need is to export figures and put them in here, with some short commentary.  
% Subsection (2) is done for heteroclinic, but I need Arnd to confirm what I'm seeing now with regards to homoclinic continuation; otherwise, its also just a matter of putting in graphs.  Subsection (3) will take a bit more work; though I have the code, there is still the question of what to present: do I show what happens if I add noise to these steady state solutions?  How do I convey that they are stable?  This aspect of the project may need to be cut out, and instead we'd just say ``direct simulations show that they are stable.''

% After this, we'll need to include some conclusion, but I'll have to talk more with Quinton about that.

\subsection{Heteroclinic Continuation in the Limit $\gamma \to 0$}
\label{section:het_cont}

\begin{figure}[h!]
\centering{\includegraphics[width=.7\columnwidth]{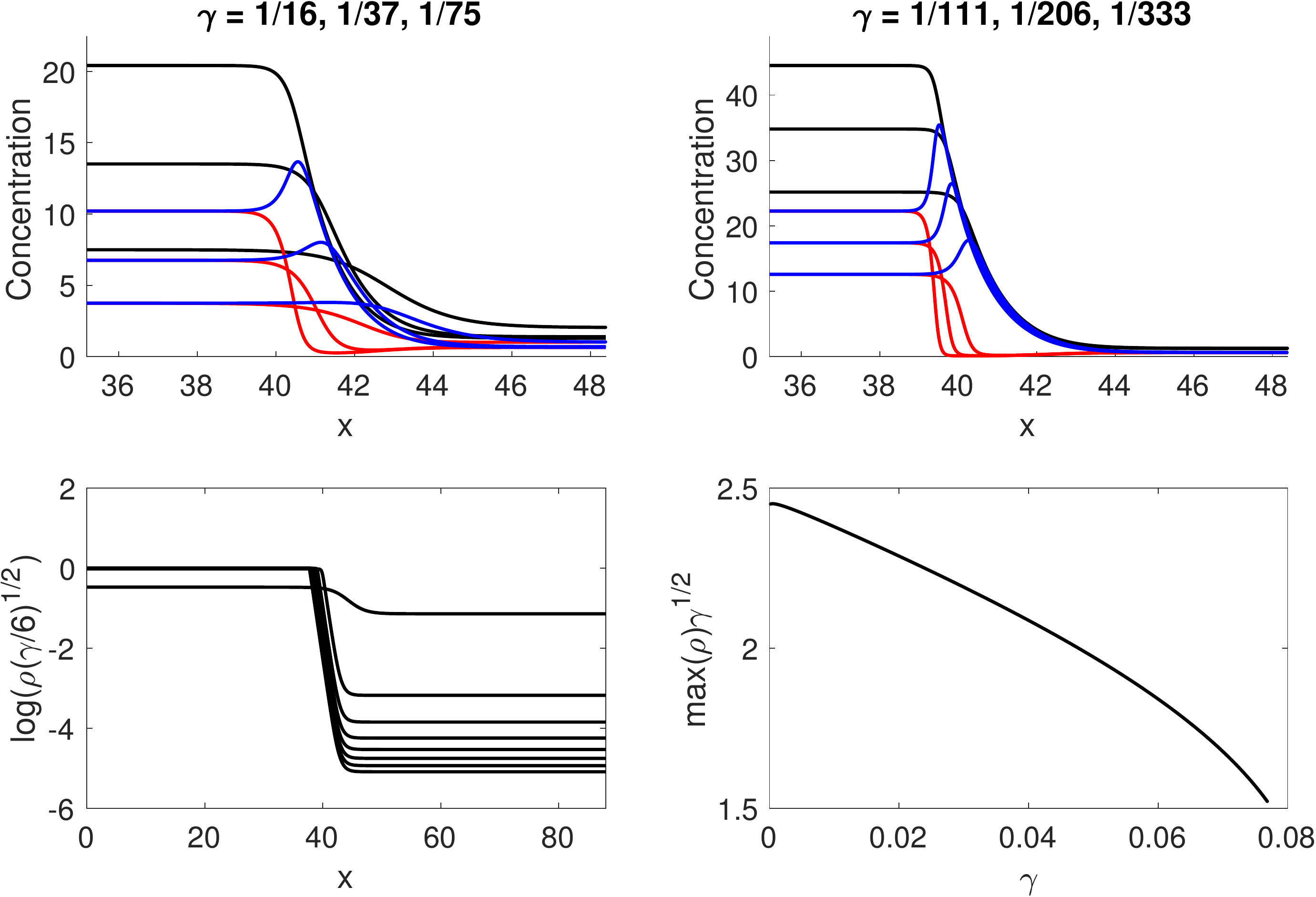}}
\caption{Heteroclinic profiles plotted as $\gamma$ varies from $\gamma = 1/13$ to $\gamma =1/6639$. Plots of $\rho\sqrt{\gamma/6}$,  $\log(\rho\sqrt{\gamma/6})$, and $\rho'/\rho$  exhibit the asymptotically simple structure of the heteroclinic. Bottom left shows the actual computational domain, grid size is $dx=0.088$.}
\label{het_conts}
\end{figure}

From Section \ref{section:asymmetric}, we know that heteroclinics exist for $\gamma$ less than and sufficiently close to $\gamma^*$.  In the present section, we present numerical evidence that heteroclinic orbits exist for all $0<\gamma<1/12$ and investigate in particular the limit $\gamma \to 0$.  The results are illustrated in  Figure \ref{het_conts}. As $\gamma$ decreases, the amplitude of the upper limit increases approximately as $\sqrt{6/\gamma}$. We also illustrate how the solution is patched using a long intermediate given by a simple exponential. We suspect that a singular perturbation analysis would reveal more details of the structure of heteroclinics at small $\gamma$. One can for instance use the scaling  $\rho = \sqrt{\gamma}\tilde\rho$ and formally set $\gamma = 0$, to obtain  a differential equation that admits exponential decay or growth solutions (i.e. $(\rho')^2 =\rho^2$).    Our numerical calculations confirm the coefficient $\sqrt{6}$ of $\gamma^{-1/2}$ in the asymptotics to an accuracy of $10^{-5}$.

% produced with more_tiled_cont_plots.m and inkscape
% dx = 0.0088
% The notion that heteroclinics converge such piecewise functions under the rescaling $\rho \to \sqrt{\gamma}\rho$ is perhaps not surprising.  Under this rescaling, our equation becomes
% \begin{equation}
% {\gamma}\rho''' = \rho'({\gamma}(1+2\mu)-2\frac{4(\rho^2 - (\rho')^2) -(\rho^2-(\rho')^2)^2}{16 +8(\rho^2 + (\rho')^2)+(\rho^2 -(\rho')^2)^2}).
% \end{equation}
% Setting $\gamma =0$, we get five possible first order equations: $\rho'=0$, $\sqrt{\rho^2-4} =$ or $\pm\rho = \rho'$.  The first admits any constant non zero solution, where the third is just exponential growth or decay at rate $1$.

\section{Temporal stability --- eigenvalues and direct simulations}
\label{section:direct}
Beyond existence, a natural next question is concerned with stability of equilibrium solutions. We address this question here from two vantage points. We first investigate the spectrum of linearized operators, and then inspect time evolution of small disturbances of spatial homoclinics and heteroclinics in direct simulations.

\paragraph{Spectral stability and instability --- continuous spectra.}
We are interested in the linearization at the stationary homoclinic and heteroclinic solutions. We therefore linearize the equation  \eqref{myxo_pde} at such a solutions and look for solutions with exponential growth in time $\rme^{\lambda t}$, 
\begin{equation}
\begin{aligned}
\lambda u &=  u_{xx} + u_x - n_u(x) u + n_v(x) v,\\
\lambda v &= v_{xx} - v_x + n_u(x) u - n_v(x) v
\label{eig_prob}
\end{aligned}%\label{e:op}
\end{equation}
where we have defined $n_u(x) = \partial_1 r(u_0,v_0) - \partial_2 r(v_0,u_0)$, and $n_v(x)=\partial_1 r(v_0,u_0) - \partial_2 r(u_0,v_0)$, and $u_0$ and $v_0$ denote the stationary solution, homoclinic or heteroclinic. The right-hand side defines an elliptic operator on the real line such that it's spectrum is contained in a sector $|\Im\lambda|\leq \beta -\alpha \Re\lambda$, $\alpha,\beta>0$, when considered on spaces of bounded continuous functions or $L^p$-spaces. Alternatively, we could also consider weighted spaces, imposing exponential growth or decay at $\pm\infty$. 

The spectrum of such operators can be decomposed into isolated eigenvalues with finite multiplicity and the complement, often referred to as the essential spectrum; see for instance \cite{fs}. In fact, one readily sees that the operator is Fredholm of index 0 for values of $\lambda$ to the right of the spectrum at the linearization at spatial infinity, $x\to\pm\infty$. We therefore collect this stability information in the following lemma. 

\begin{Lemma}\label{l:ess}
 Consider the linearization at a homoclinic solution $(u_0,v_0)(x)$ \eqref{eig_prob} on $L^2(\R,\R^2)$. Then the essential spectrum of the linearization (where the operator defined in \eqref{eig_prob} is not Fredholm of index 0) is given by the curves
 \begin{equation}\label{e:ess}
  \lambda_\pm(k)=-a+k^2 \pm \sqrt{a^2-k^2},
 \end{equation}
where $a=r_u(u_\infty,v_\infty)$, $u_\infty=v_\infty$ is the asymptotic state of the homoclinic. For heteroclinic orbits with limits $(u,v)_\infty^\pm$, the essential spectrum is to the left  of the four curves defined by \eqref{e:ess} in the complex plane. In particular, stable homoclinics have $\gamma<1/8$ and 
\begin{equation}\label{e:in}
 u_\infty^2/2\not\in \left( \frac{1 -2\gamma -\sqrt{1-8\gamma}}{\gamma+\gamma^2}, \frac{1 -2\gamma +\sqrt{1-8\gamma}}{\gamma+\gamma^2}\right). 
\end{equation}
\end{Lemma}
\begin{Proof}
 By standard compactness argument, see for instance \cite{fs} and references therein, it is sufficient to compute the spectra at the asymptotic state using Fourier transform. Setting therefore $n_{u/v}=n_{u/v}^\infty$, and using Fourier transform, one is left with computing eigenvalues of a matrix
 \begin{equation}\label{e:mat}
  \lambda\left(\begin{array}{c}u\\v\end{array}\right)=\left(\begin{array}{cc} -k^2+\rmi k  +a &-a \\ -a& -k^2-\rmi k +a \end{array}\right)\left(\begin{array}{c}u\\v\end{array}\right)
 \end{equation}
where we used that $n_u^\infty=n_v^\infty=-r_u+r_v=:a$ when evaluated on  $u=v$. Computing the eigenvalues readily gives \eqref{e:ess}. Evaluating the condition $a>0$ leads to the condition on $u_\infty=v_\infty$ after some straightforward algebra.
 \end{Proof}
\begin{Corollary}\label{c:inst}
 Small clusters and gaps are spectrally unstable, that is, there is $\delta>0$ such that clusters and gaps with background states in $(\rho_--\delta,\rho_-)$ and $(\rho_+,\rho_++\delta)$ are spectrally unstable. 
\end{Corollary}
\begin{Proof}
 It is sufficient to note that $u=\rho_-/2$ lies in the interval identified in \eqref{e:in}.
\end{Proof}
One can rephrase this result from a bifurcation theory perspective. Clusters and gaps bifurcate from the homogeneous background state. This bifurcation does however not occur when the background state changes stability in the sense that the essential spectrum crosses the imaginary axis. This instability occurs \emph{before} the bifurcation of homoclinics, such that all homoclinics are unstable against perturbations of the background state. Since bifurcation typically is associated with a degeneracy in the kernel of the linearization, one may wonder what precisely happens at the origin on a spectral level, that gives rise to the bifurcation of localized clusters and gaps. 

A slightly weaker spectral stability criterion refers to the possibility of, roughly speaking,  choosing optimal exponential weights. The resulting ``optimized spectrum'' was coined the ``absolute spectrum'' in \cite{ssabs} and shown to generically be the limit of spectra in bounded domains as the domain size goes to infinity. In order to find the absolute spectrum, one computes the complex values of $k$ that are compatible with a fixed value of $\lambda$. ordering the resulting four roots by $\Re\nu$,  $\nu=\rmi k$, that is, by the imaginary part of $k$, one looks for values of $\lambda$ such that $\Re\nu_1<\Re\nu_2=\Re\nu_3\leq \Re\nu_4$. It turns out that, by reversibility, essential spectra and absolute spectra coincide; see also \cite{rss}. 

Yet more refined stability criteria are concerned with pointwise stability of the linear evolution. Here, one constructs the temporal evolution near the states at $\pm\infty$ exploiting translational invariance and studies decay properties of the temporal Green's function, that is, of solutions with Dirac initial data; see \cite{holzerscheel}. Typically, such instabilities correspond to pinched double roots crossing the imaginary axis. The following lemma states, roughly speaking, that the bifurcation of homoclinic orbits does not correspond to an onset of a pointwise instability, but rather to a secondary instability due to a degenerate pinched double root.  

\begin{Lemma}\label{l:pdr}
 For all $a>0$, there exists a simple unstable pinched double root at $\lambda=2a$, such that the system is pointwise unstable whenever the essential spectrum is unstable. At $\rho_\infty=\rho_\pm$, or $a=1/2$, a double root $\lambda=0,\nu=0$ is degenerate. 
\end{Lemma}
\begin{Proof}
 Following \cite{holzerscheel}, we compute the complex dispersion relation by evaluating the characteristic polynomial in \eqref{e:mat} with $\rmi k=\nu$ to find
 \[
 d(\lambda,\nu)= \lambda^2-2(a+\nu^2)\lambda +\nu^2(\nu^2+2a-1)=0.
 \]
 Solving for double roots $d=\partial_\nu d=0$, we find the four solutions $(\lambda,\nu)=(0,0),(2a,0),((a-1/2)^2,\pm\sqrt{a^2-1/4})$. Since $\partial_\lambda d\neq 0$ at the double root $\lambda=2a$, it is non-degenerate and creates a pointwise instability for $a>0$. 
\end{Proof}

\paragraph{Spectral stability --- eigenvalues of heteroclinics.}
We did not attempt to study eigenvalues of the linearization \eqref{eig_prob} analytically. The discussion of essential spectra, above, indicates that in the most accessible limits of small amplitude, homoclinic orbits possess unstable essential spectrum and are, even in a very weak sense, unstable due to the instability of the background state. In the following, we present numerical results. We substituted the numerically computed profiles $(u_0,v_0)$ into a second-order upwind discretization of \eqref{eig_prob}. We introduced artificial boundary conditions $u'(0) = v(0) = 0$ on the left and $u(L) = v'(L) = 0$, roughly imposing zero values for populations migrating inwards and zero derivative for outward moving populations. 

\begin{figure}[H]
\centering
\includegraphics[width = .49\columnwidth]{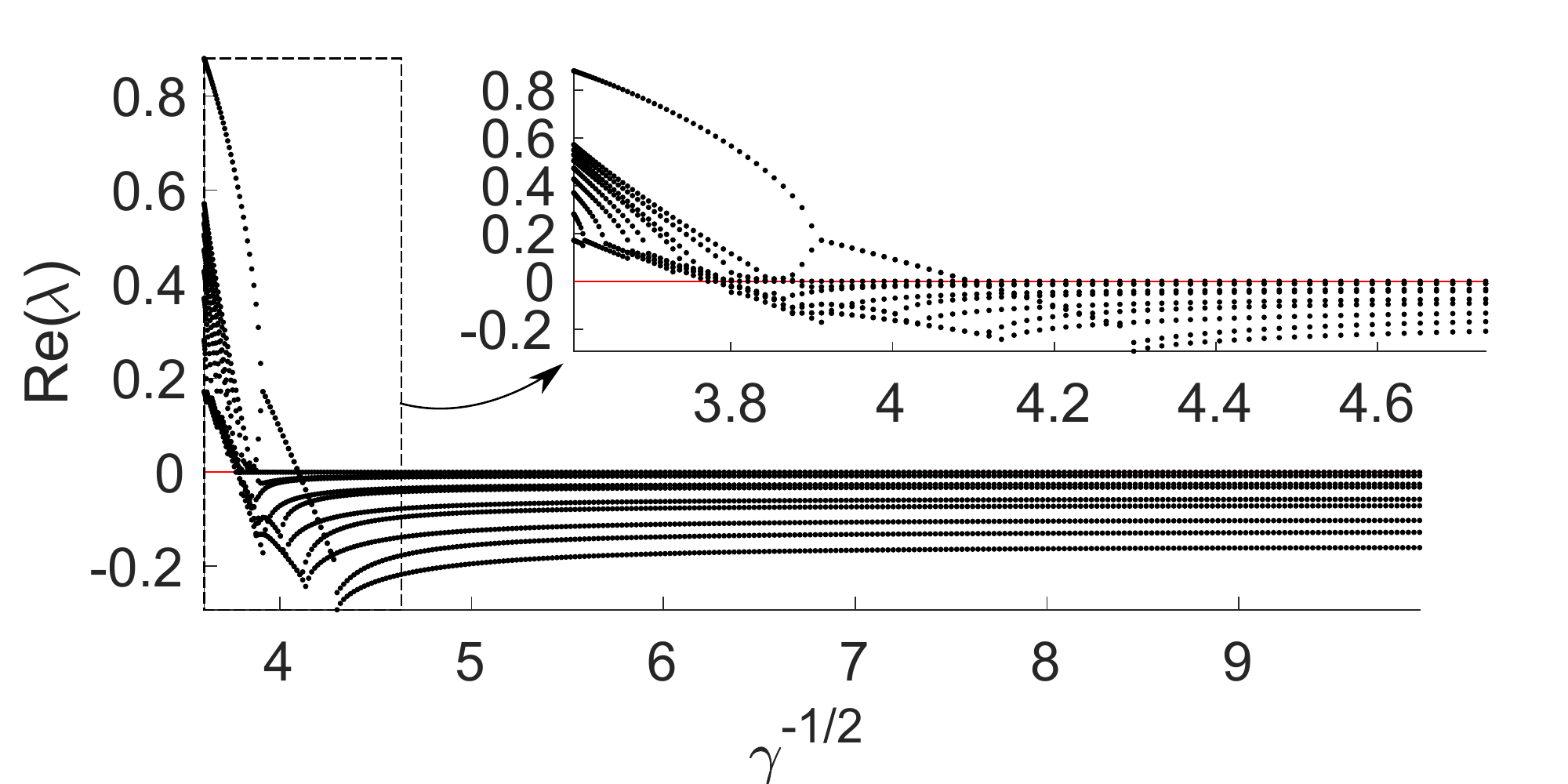}
\hspace{1cm}
\includegraphics[width = .21\columnwidth]{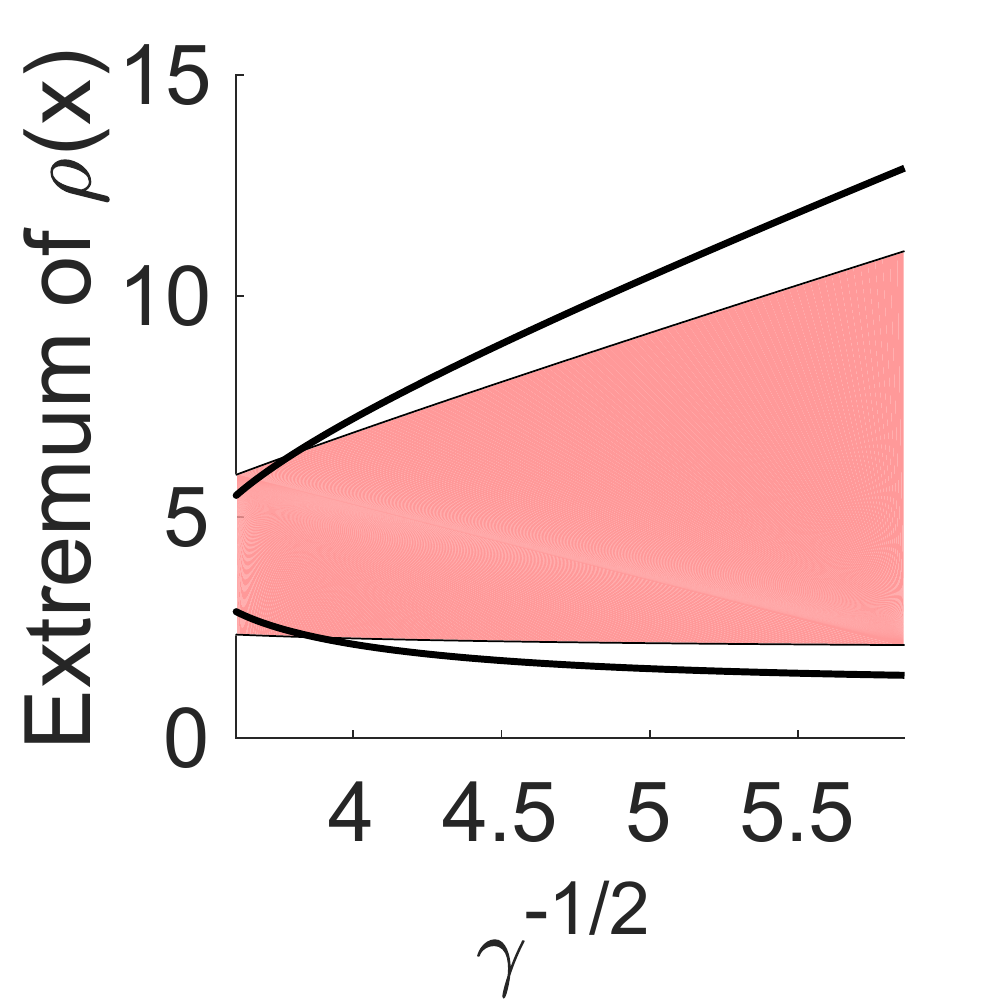}
\caption{Left we have the real part of the spectrum of the heteroclinics.  Right, we have the two background concentrations of the heteroclinics in black (numerically the maximum and minimum of $\rho(x)$, and the region where the corresponding constant solutions are stable in pink. Computations here use grid size from the  previous heteroclinic continuation.}
 \label{hetstab}
\end{figure}

We first present results for heteroclinic orbits. We used Matlab's sparse eigenvalue solver to find the 10 eigenvalues nearest to 1 for the discretized problem. Results appear to robustly give the 10 leading eigenvalues when other preconditioners for the eigenvalue are employed. Figure \ref{hetstab} shows that heteroclinic orbits near the bifurcation point $\gamma=\gamma_*=1/12$ are unstable as expected. For $\gamma\sim 1/16$, the limit states of the heteroclinic stabilize and we see stability for all values of $\gamma$ less than this critical value. Figure \ref{hetstab} shows that background states stabilize at about $1/\sqrt{\gamma}=3.8\ldots$, at which point eigenvalue clusters stabilize. We note however that heteroclinics stabilize somewhat later, when a pair of isolated eigenvalue crosses the imaginary axis. 
% made with graph_het.m, graph_bgsts.m
%dx = 0.0088
% \pagebreak
%% made with 
%\begin{wrapfigure}{l}{0.5\columnwidth}
%\def\svgwidth{0.5\columnwidth}
%\input{figures/bgst_het_cont.pdf_tex}
%\caption{Comparison of background states $\rho_\pm^*$ and roots $\rho^0_\pm$.}
%\end{wrapfigure} 
We tested stability and instability in direct simulations with reasonably good agreement. Some results are shown below.

% dx = 5e-3, domain same as in continuations

\begin{figure}[H]
\centering {
\includegraphics[width=.7\columnwidth]{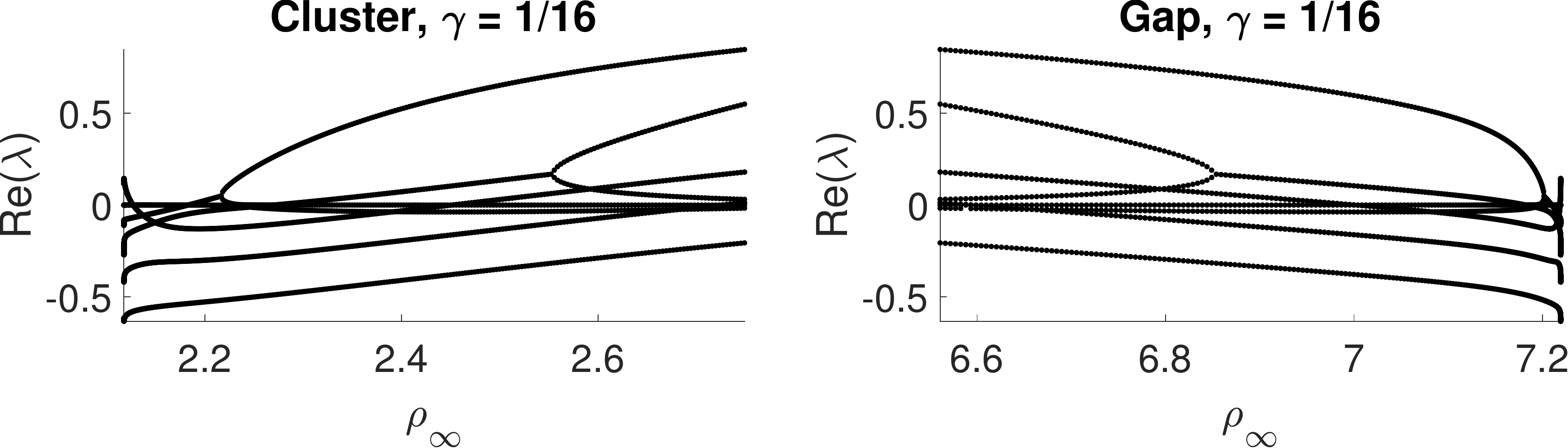}
\includegraphics[width=.7\columnwidth]{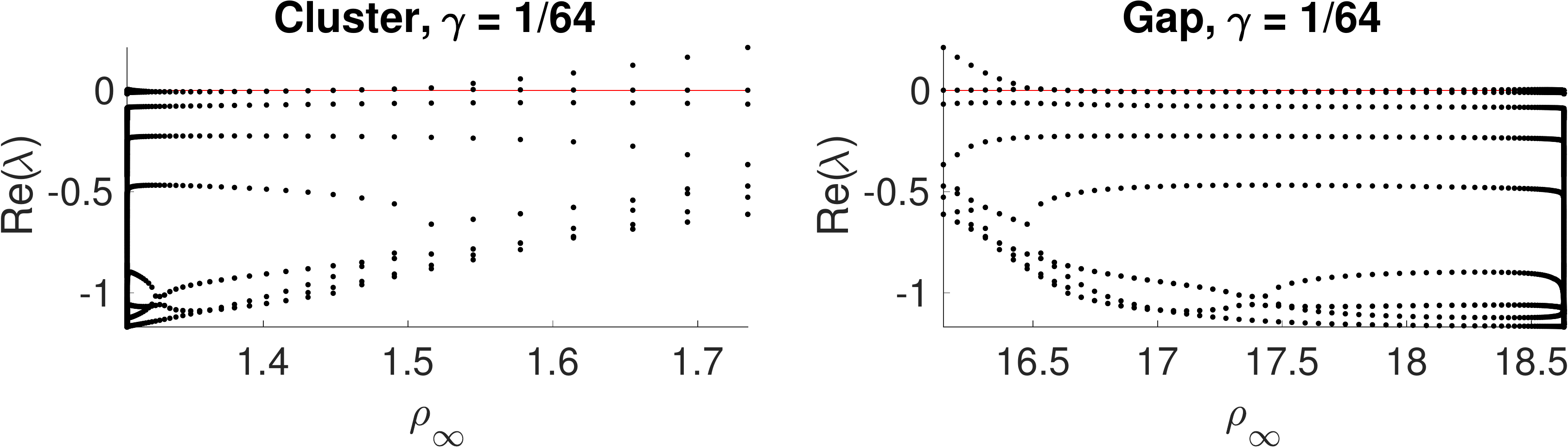}
\includegraphics[width=.7\columnwidth]{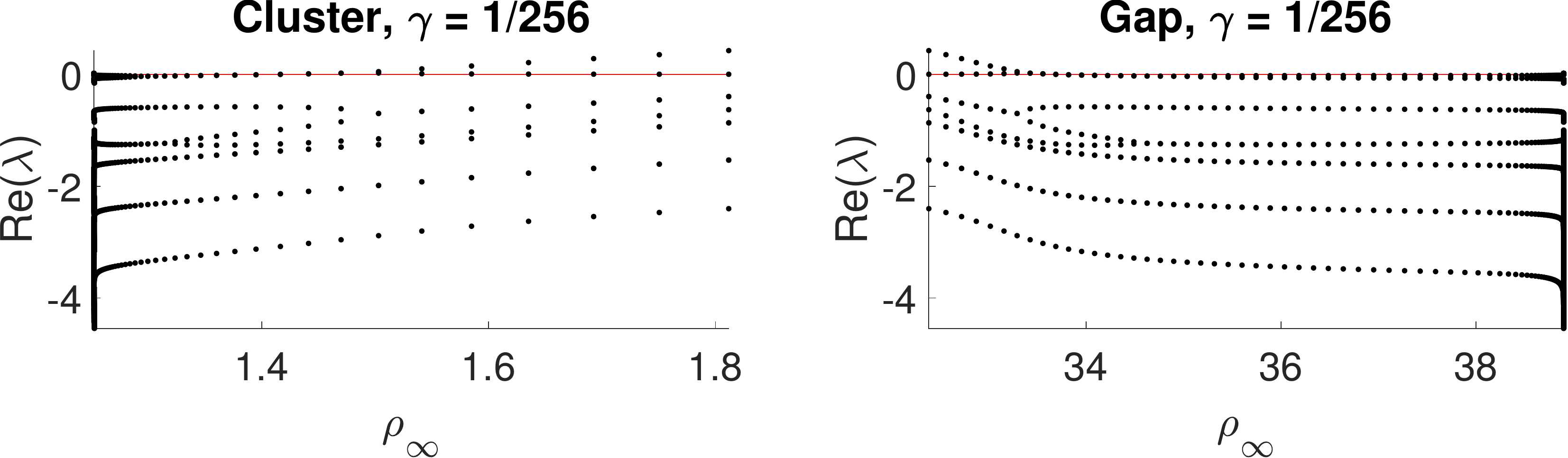}}
\caption{Spectra of clusters and gaps as functions of the background $\rho_\infty$, for various values of $\gamma$. Note that eigenvalues with positive real parts exist for  gaps and clusters with large or small $\rho_\infty$, that is, near small-amplitude or heteroclinic limit, respectively.}
\label{bgst_eigs}
\end{figure}

\paragraph{Spectral stability --- eigenvalues of homoclinics.}
We next turn to homoclinics, clusters and gaps. We focus on small values of $\gamma$, where the limiting heteroclinic orbit is stable. Figure \ref{bgst_eigs} shows largest eigenvalues of clusters and gaps, depending on the background state $\rho_\infty$, for various values of $\gamma$. Both gaps and clusters are unstable  in the small-amplitude as well as in the heteroclinic limit, but appear to be stable for an intermediate range of values of $\rho_\infty$, which grows as $\gamma$ decreases. We next show direct simulations that confirm these spectral computations while also illustrating the nature of the instability. 

\begin{figure}[H]
\includegraphics[width=.49\columnwidth]{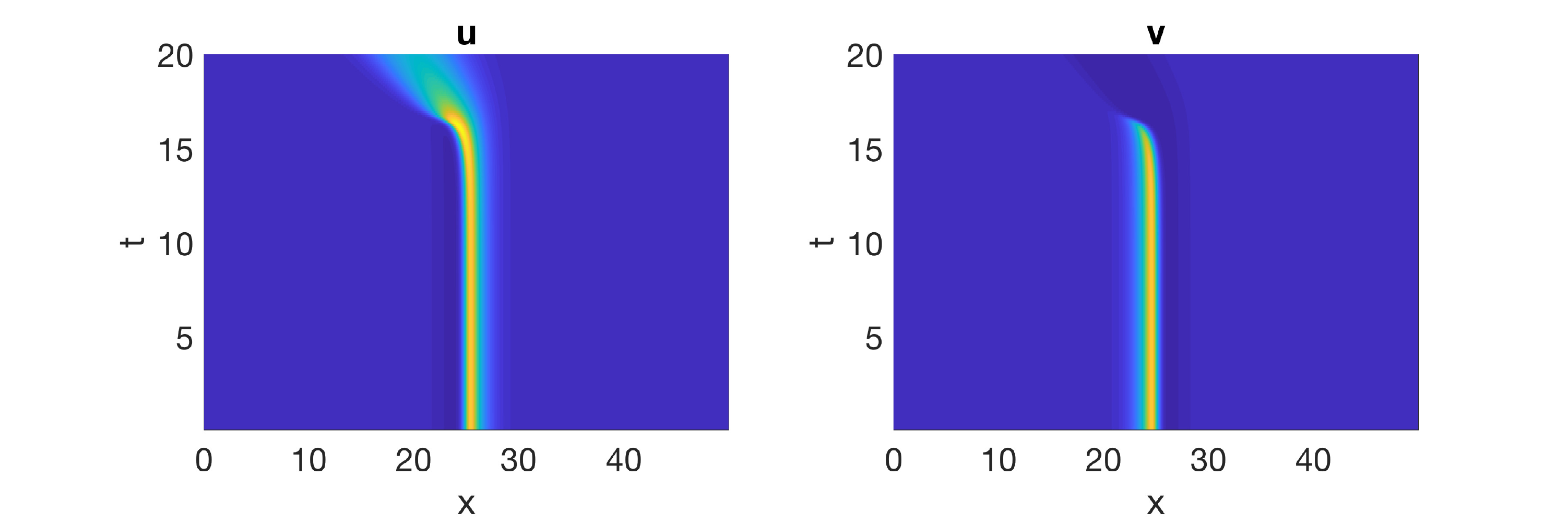}\hfill \includegraphics[width=.49\columnwidth]{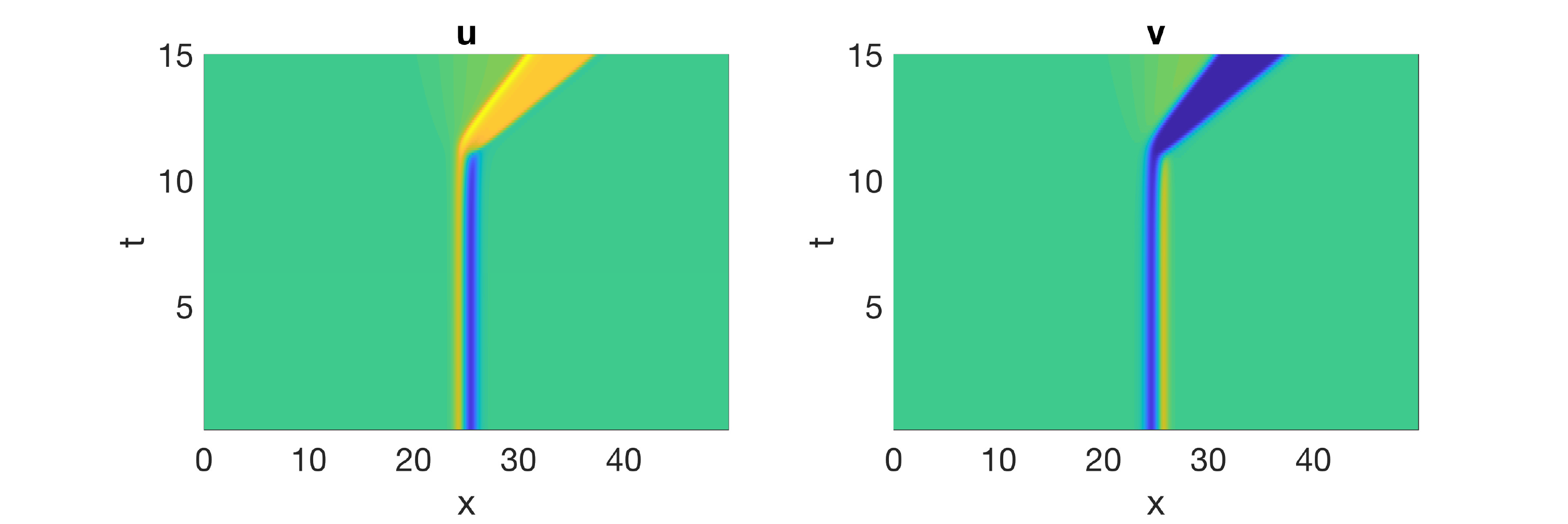}\\
\includegraphics[width=.46\columnwidth]{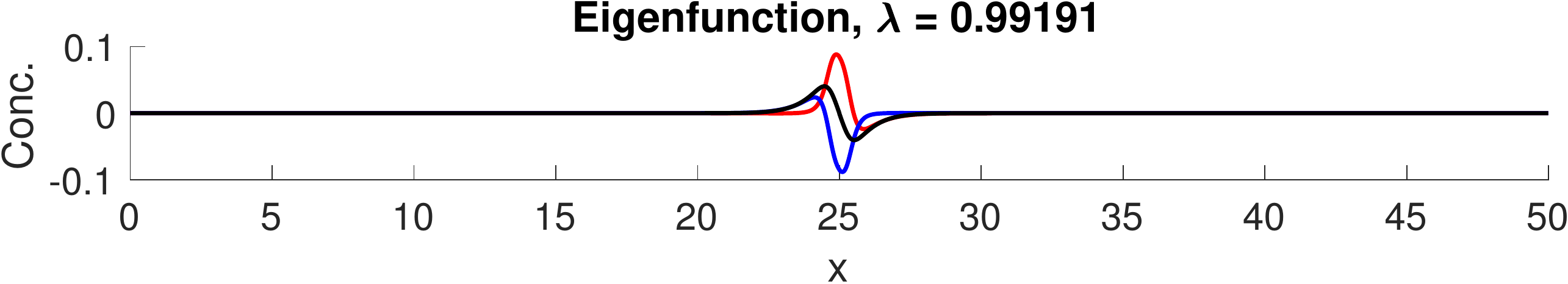}\hfill
\includegraphics[width=.46\columnwidth]{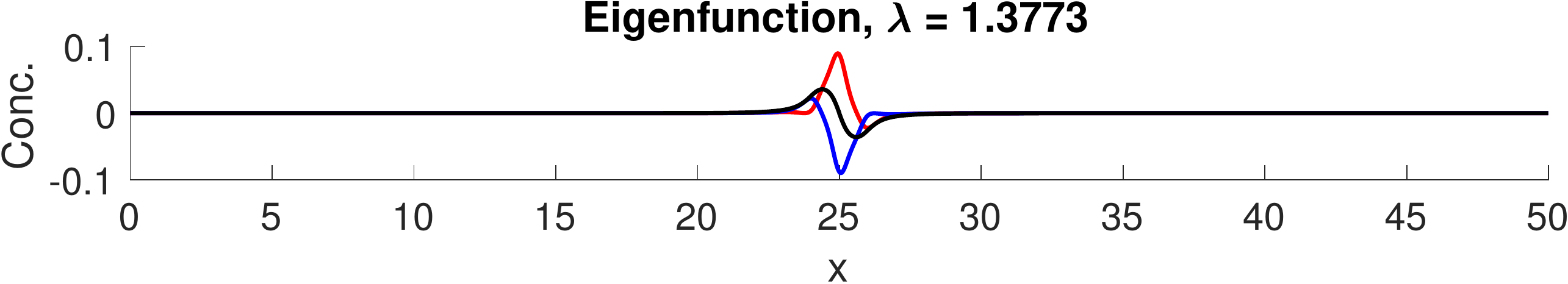}\\
\includegraphics[width=.46\columnwidth]{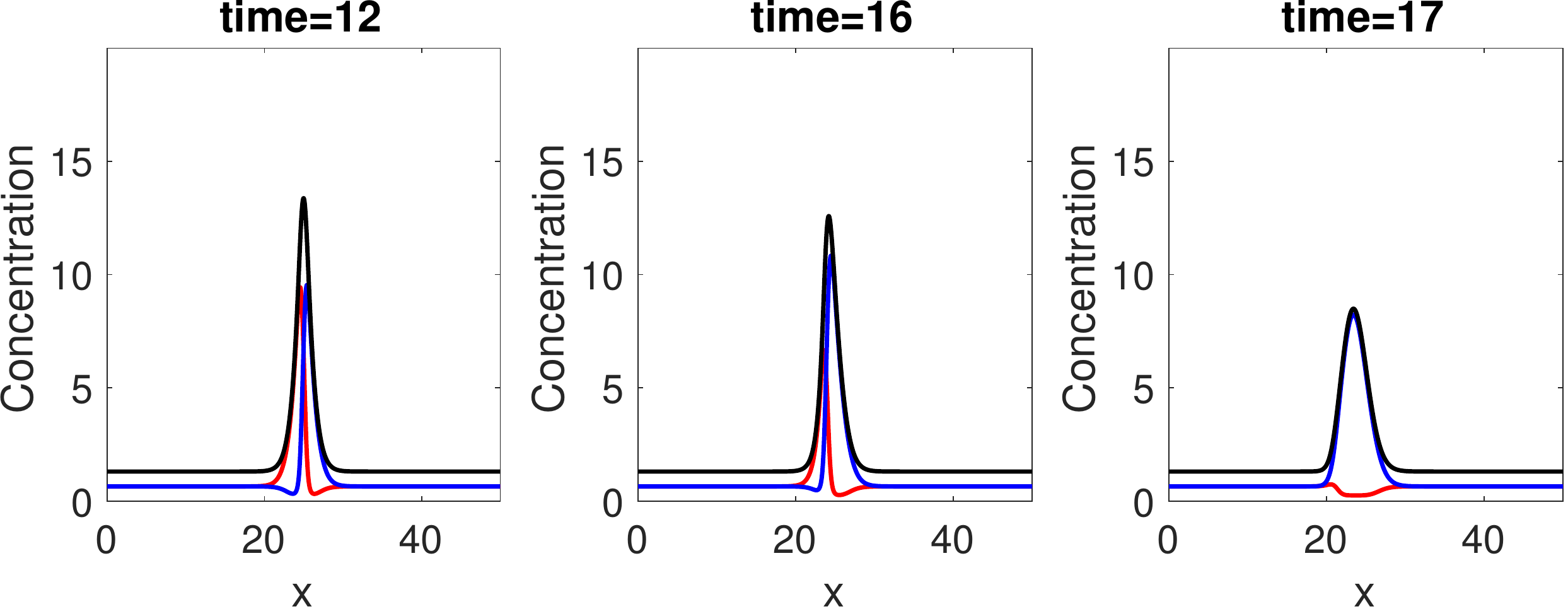}\hfill
\includegraphics[width=.46\columnwidth]{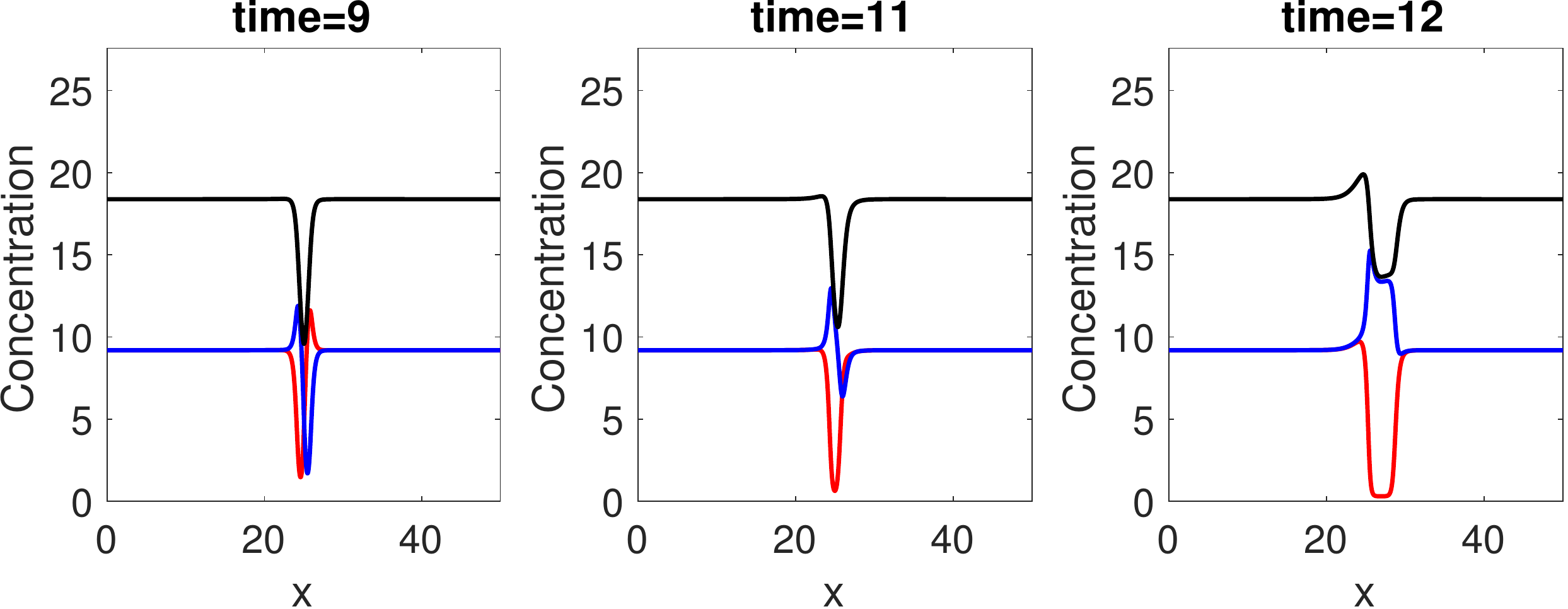}
\caption{Cluster instability (left) and gap instability (right). Time evolution of perturbation of a stationary profile in the direction of the unstable eigenvector. Shown are space-time plots  for $u$ and $v$ (top row), shape of the most unstable eigenfunction (middle row), and snap shots of time evolution for $u$ (red), $v$ (blue), and $u+v$ (black). Parameter values are  $\gamma = 1/64$, $\rho_\infty=1.3115$ (left) and $\rho_\infty=18.3805$ (right).}\label{f:gc}
\end{figure}

\paragraph{Direct simulations.} 
We performed direct simulations using a second order upwind finite-difference method with grid size $5\cdot10^{-3}$ and Matlab's stiff solver \textsc{ode15s}. As initial conditions, we used the steady state with a small perturbation in the direction of the leading eigenvector. Evolution of instabilities of clusters and gaps is shown in Figure \ref{f:gc}, showing a drift instability followed by diffusive decay. We observed different types of instabilities, in particular for moderate values of $\gamma$, without finding simple organizing principles, there. 

% 
% Need figure showing:
% 
% - stability of heteroclinics\\
% - weak instability due to point eigenvalue, oscillatory, of heteroclinics\\
% - instability of a gap, only one example, maybe when the gap is close to the heteroclinic limit\\
% - instability of cluster in regime near heteroclinic\\
% - stability of cluster in the intermediate regime\\
% 
% all figures, maybe complement (as you did) plot of u,v,rho at some fixed time, and a space-time plot of u and v

% dx = 5e-3 
% 1e-4 L^2 norm eigenvector perturbation -- edit: actually I forgot the factor of sqrt(dx) in the normalization, so this perturbation is more like 1e-4*sqrt(dx) = 7e-6 (in L^2, as opposed to l^2)

% dx = 5e-3
% % 1e-3 L^2 norm eigenvector perturbation -- edit: see comment above, eig pert actually 7e-5
% \begin{figure}[H]
% \centering {}
% \caption{gap instability, $\gamma = 1/64$.}
% \end{figure}

% 
% 
% \begin{figure}[H]
% \centering {
% \includegraphics[width=.66\columnwidth]{lower_noise0_eig0.0001_gam64_pcolor.pdf}
% \includegraphics[width=.6\columnwidth]{lower_noise0_eig0.0001_gam64_eigenfunction.pdf}
% \includegraphics[width=.6\columnwidth]{lower_noise0_eig0.0001_gam64_snapshot.pdf}}
% \caption{cluster instability, $\gamma = 1/64$.}
% \end{figure}
% 
% % dx = 5e-3
% % 1e-3 L^2 norm eigenvector perturbation -- edit: see comment above, eig pert actually 7e-5
% \begin{figure}[H]
% \centering {
% \includegraphics[width=.66\columnwidth]{upper_noise0_eig0.001_gam64_pcolor.pdf}
% \includegraphics[width=.6\columnwidth]{upper_noise0_eig0.001_gam64_eigenfunction.pdf}
% \includegraphics[width=.6\columnwidth]{upper_noise0_eig0.001_gam64_snapshot.pdf}}
% \caption{gap instability, $\gamma = 1/64$.}
% \end{figure}

% dx = 4.4e-3
% eig pert 1e-3 -- edit: see comment above; eig pert 6.6e-5

\begin{figure}[h!]
\begin{minipage}{0.5\textwidth}
\includegraphics[width=\columnwidth]{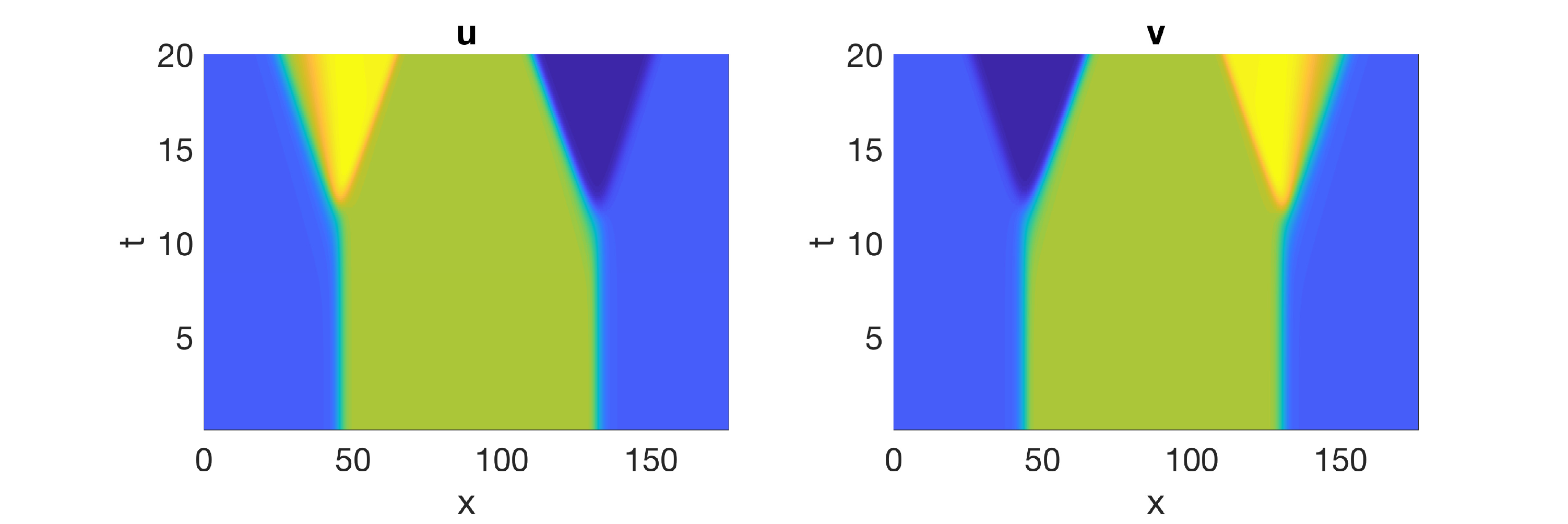}\\[0.2in]
\includegraphics[width=.95\columnwidth]{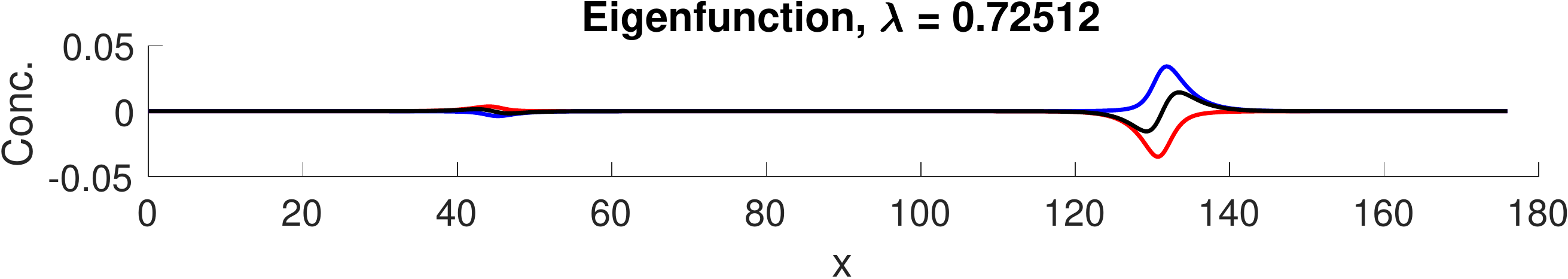}\\[0.2in]
\includegraphics[width=.95\columnwidth]{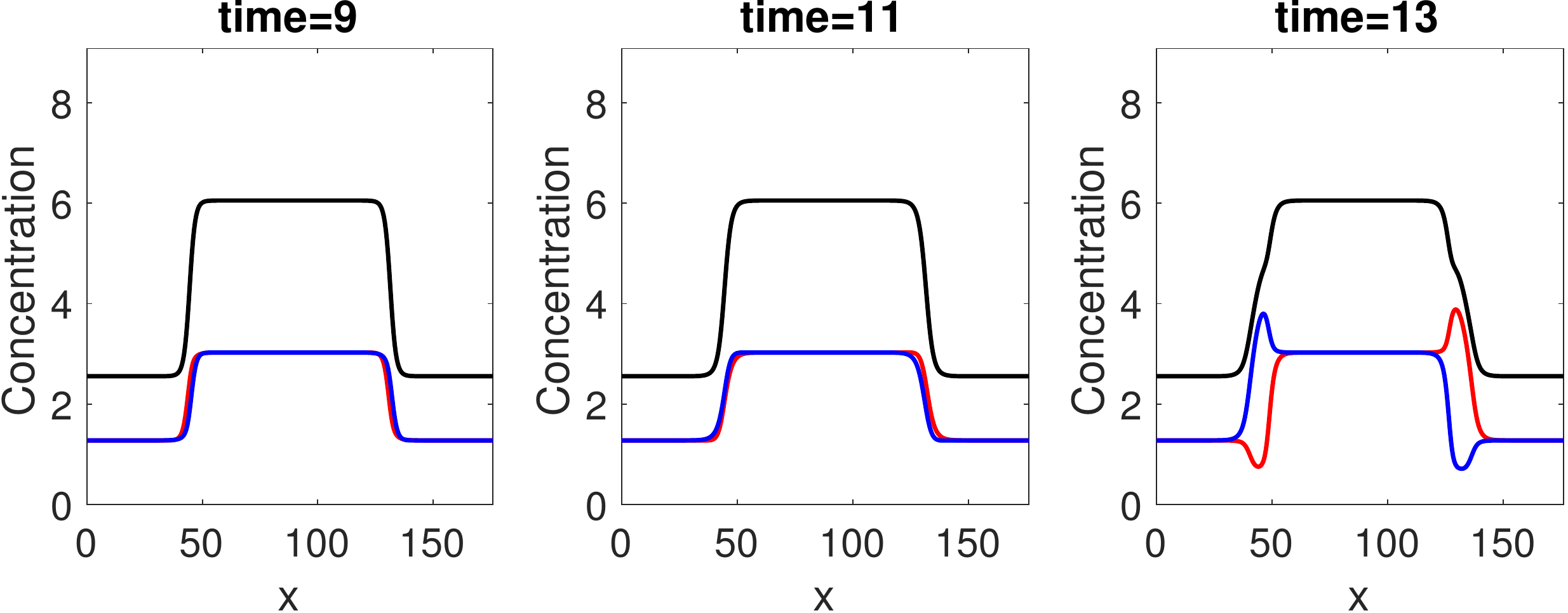}
\end{minipage}\hfill 
\begin{minipage}{0.48\textwidth}
\includegraphics[width=\columnwidth]{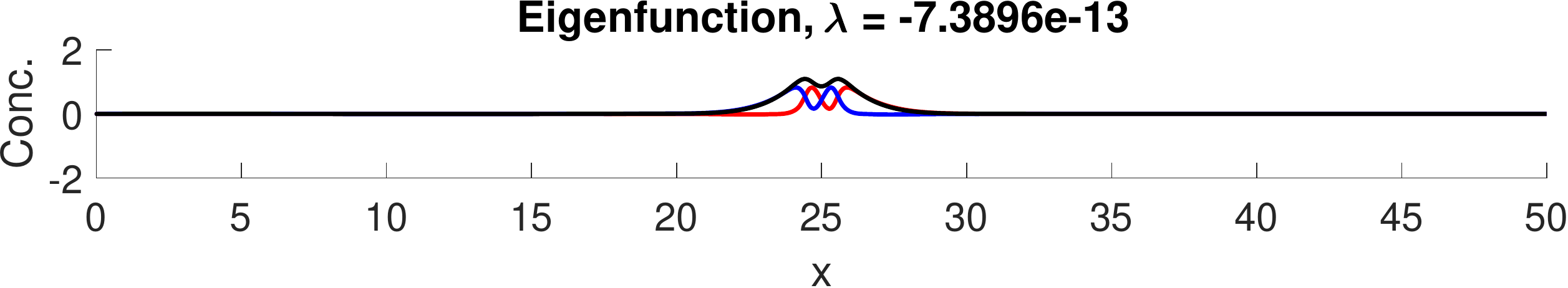}\\
\includegraphics[width=.33\columnwidth]{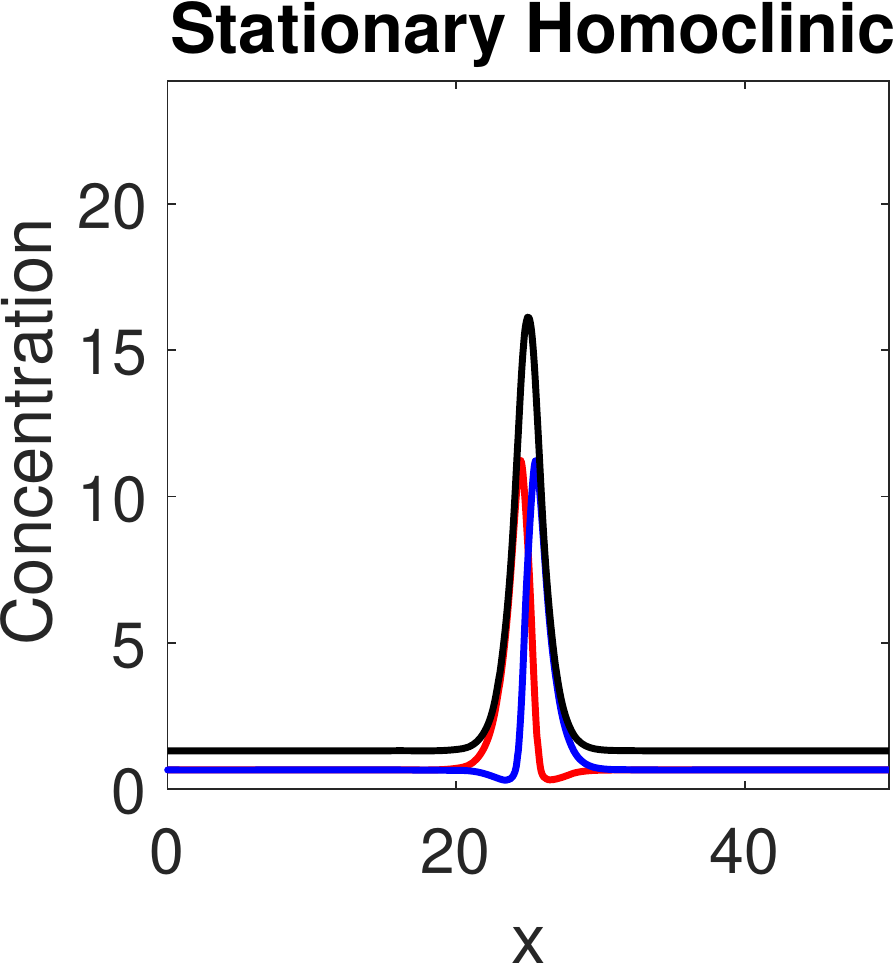}
\includegraphics[width=.66\columnwidth]{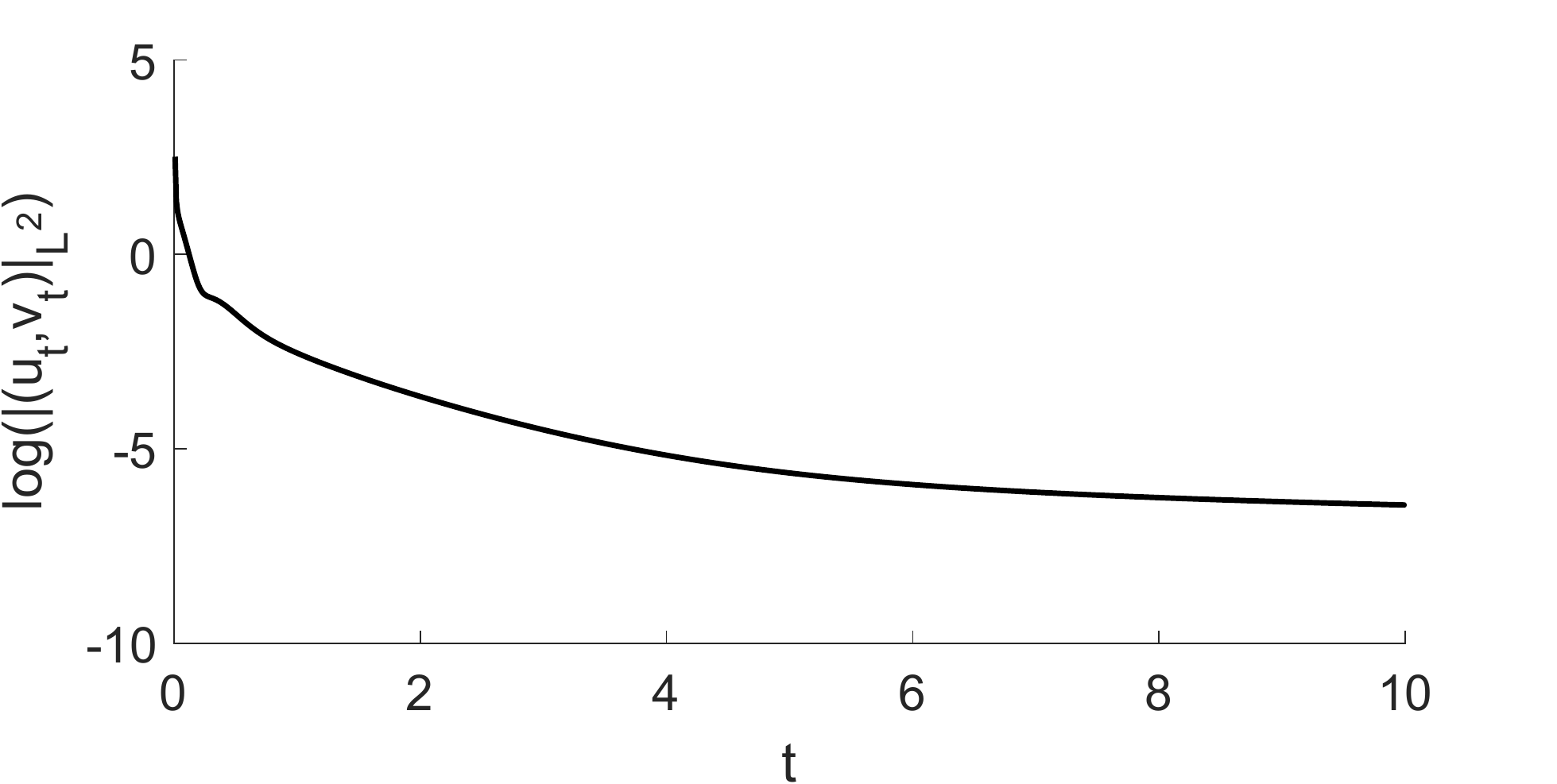}
\includegraphics[width=\columnwidth]{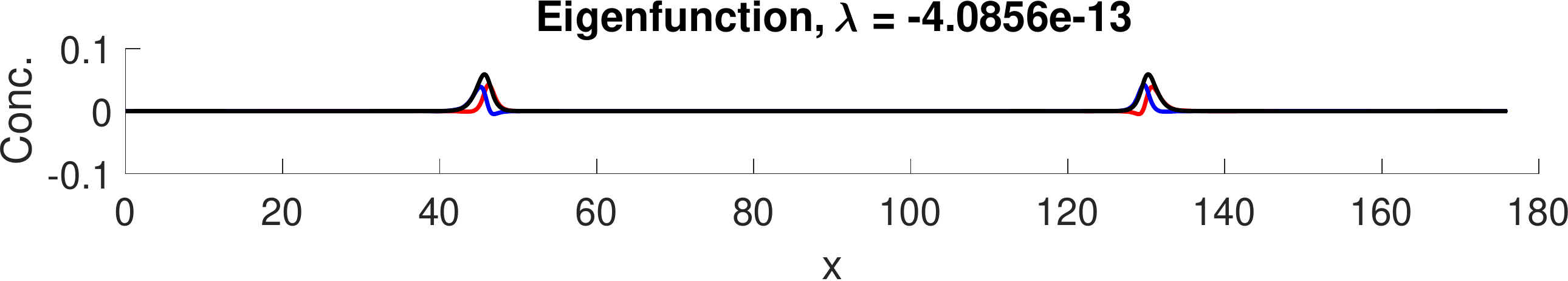}\\
\includegraphics[width=.33\columnwidth]{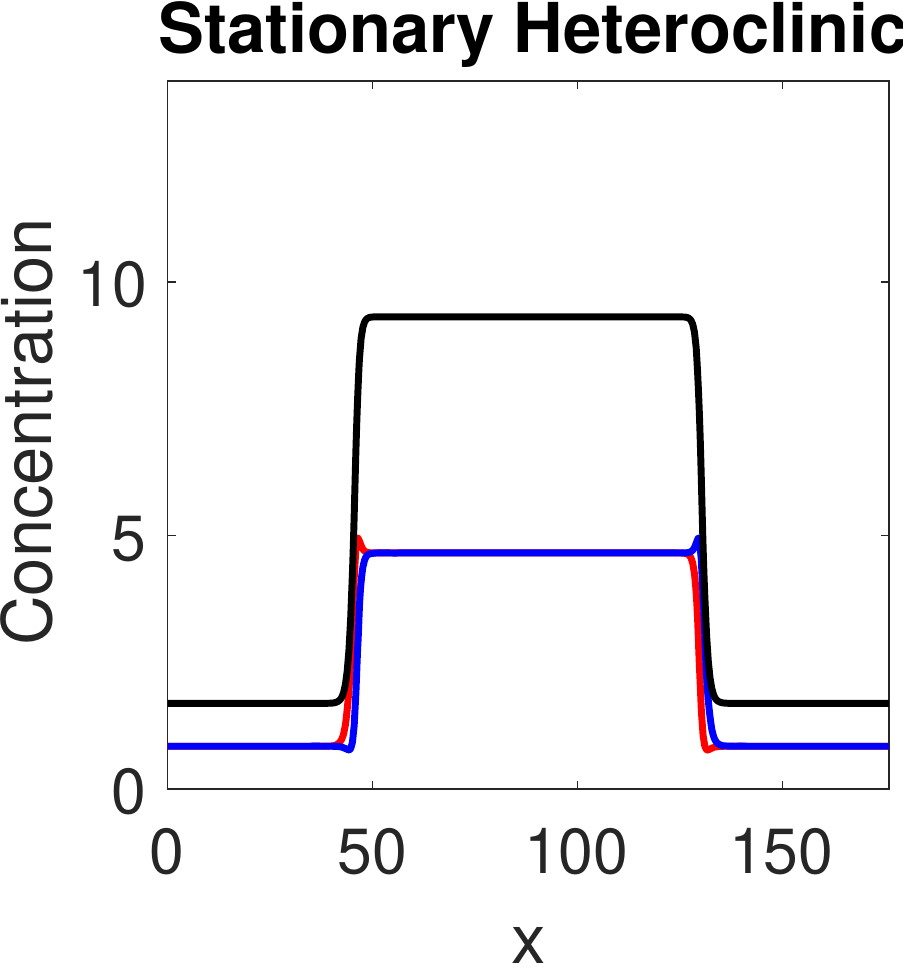}
\includegraphics[width=.66\columnwidth]{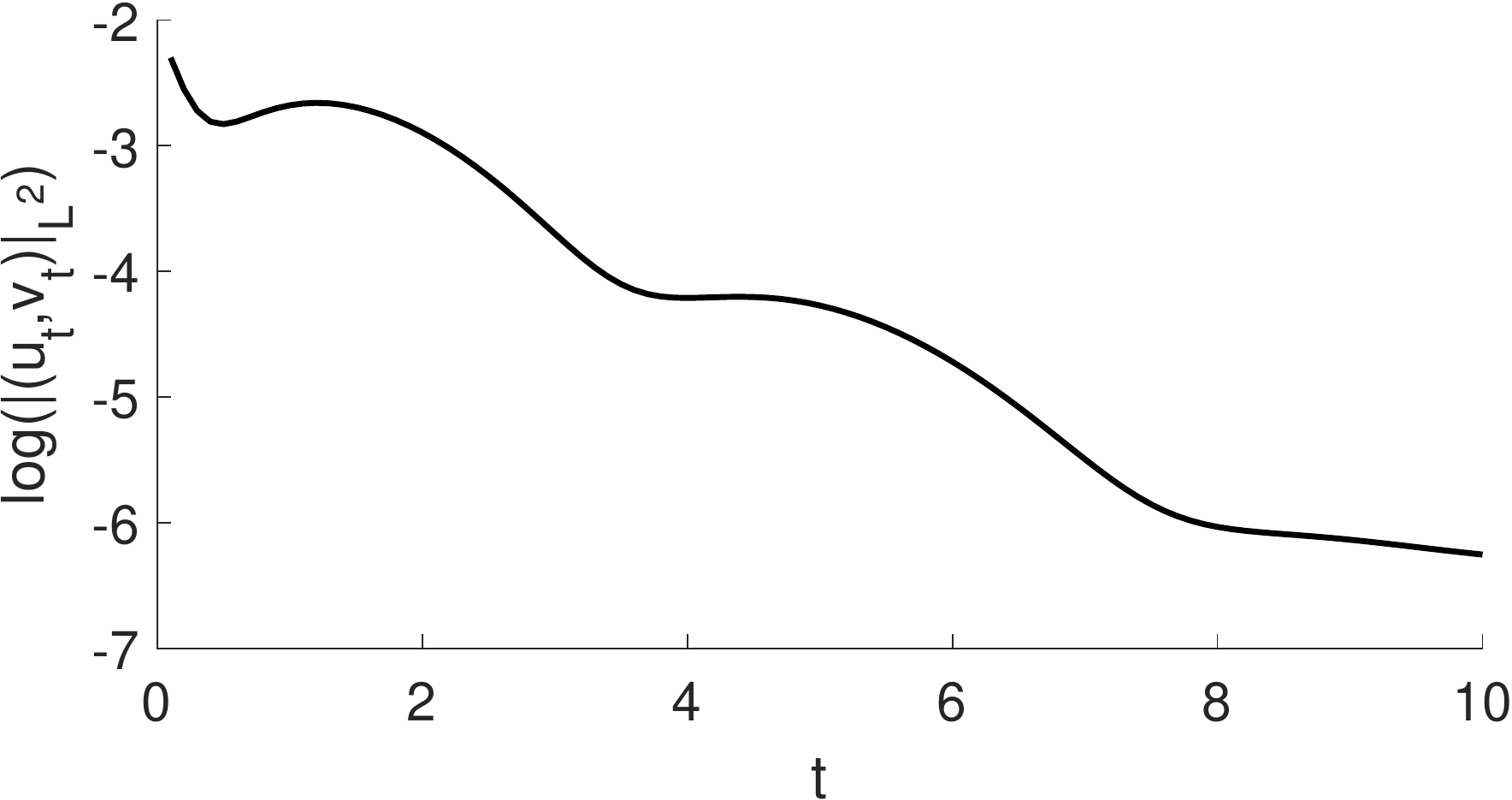}
\end{minipage}

\caption{Instability of cluster boundaries (left), showing space-time plots for $u$ and $v$, top row, profile of the leading eigenfunction, and time snapshots, for $\gamma = 1/13.81$. Stable clusters ($\gamma=1/64$, $\rho=1.3072$) and cluster  boundaries ($\gamma=1/21.41$) on the right, with leading eigenfunction corresponding to mass change (cluster) and translation (cluster boundary), respectively. }\label{f:hss}
\end{figure}

Figure \ref{f:hss} shows instability of heteroclinic profiles for $\gamma=1/13.81$. Disturbances lead to a break-up of cluster boundaries through splitting into left- and right-traveling packets. Also shown in Figure \ref{f:hss} are results of simulations in the stable parameter regime, for both clusters and cluster  boundaries. Leading eigenvalues in the stable regime were $\lambda=0$ within numerical accuracy. Decay was roughly exponential in the case of cluster boundaries. For clusters, we chose parameter values very close to the critical value and observed the expected slow decay. Note that these particular simulations, $\gamma=1/64$, $\rho_\infty=1.3115$ in Figure \ref{f:gc}, left,  $\rho_\infty=1.3072$, Figure \ref{f:hss}, top right, confirm with good accuracy the results from the spectral calculations Figure \ref{bgst_eigs}, where we found stability change at approximately $\gamma=1.3073$. Similarly, Figure \ref{f:gc}, right, demonstrates a weak instability at $1/\gamma=18.3805$, slightly below the instability threshold $1/\gamma=18.6307$ found in  \ref{bgst_eigs}, middle right.

\section{Discussion} \label{section:discussion}

We exhibited how a very simple mechanism, running and tumbling with a density-dependent, monotonically increasing tumbling rate, can lead to the self-organized formation of clusters. The analysis was based on the study of ordinary differential equations, exhibiting homoclinic and heteroclinic solutions. We also presented ample numerical evidence that some, but not all of these solutions are stable in the PDE. 

There are clearly many open questions, ranging from the study of traveling (rather than standing) clusters \cite{fuhrmann}, over a more analytical approach to stability using for instance Evans function techniques, and nonlinear stability questions. 

Beyond the immediate interest of these peculiar structures, one may be interested in more global descriptions of the dynamics of run-and-tumble processes, such as the possibility of blowup in finite time, the formation of rippling patterns, or conditions for equidistribution of agents in space and between left- and right-traveling populations; see \cite{scheel2016wavenumber,kss}. 
In this context, it would be interesting to study the role of the structures found here in the inviscid limit, $\eps\to 0$, their relation to rippling patterns studied in \cite{scheel2016wavenumber}, and to blowup studied in \cite{kss}.

\end{document}

%% file: phase_homs.pdf_tex
%% Creator: Inkscape inkscape 0.92.2, www.inkscape.org
%% PDF/EPS/PS + LaTeX output extension by Johan Engelen, 2010
%% Accompanies image file 'phase_homs.pdf' (pdf, eps, ps)
%%
%% To include the image in your LaTeX document, write
%%   \input{<filename>.pdf_tex}
%%  instead of
%%   \includegraphics{<filename>.pdf}
%% To scale the image, write
%%   \def\svgwidth{<desired width>}
%%   \input{<filename>.pdf_tex}
%%  instead of
%%   \includegraphics[width=<desired width>]{<filename>.pdf}
%%
%% Images with a different path to the parent latex file can
%% be accessed with the `import' package (which may need to be
%% installed) using
%%   \usepackage{import}
%% in the preamble, and then including the image with
%%   \import{<path to file>}{<filename>.pdf_tex}
%% Alternatively, one can specify
%%   \graphicspath{{<path to file>/}}
%% 
%% For more information, please see info/svg-inkscape on CTAN:
%%   http://tug.ctan.org/tex-archive/info/svg-inkscape
%%
\begingroup%
  \makeatletter%
  \providecommand\color[2][]{%
    \errmessage{(Inkscape) Color is used for the text in Inkscape, but the package 'color.sty' is not loaded}%
    \renewcommand\color[2][]{}%
  }%
  \providecommand\transparent[1]{%
    \errmessage{(Inkscape) Transparency is used (non-zero) for the text in Inkscape, but the package 'transparent.sty' is not loaded}%
    \renewcommand\transparent[1]{}%
  }%
  \providecommand\rotatebox[2]{#2}%
  \ifx\svgwidth\undefined%
    \setlength{\unitlength}{1152bp}%
    \ifx\svgscale\undefined%
      \relax%
    \else%
      \setlength{\unitlength}{\unitlength * \real{\svgscale}}%
    \fi%
  \else%
    \setlength{\unitlength}{\svgwidth}%
  \fi%
  \global\let\svgwidth\undefined%
  \global\let\svgscale\undefined%
  \makeatother%
  \begin{picture}(1,0.25)%
    \put(0,0){\includegraphics[width=\unitlength,page=1]{phase_homs.pdf}}%
    \put(0.25873001,0.21760983){\color[rgb]{0,0,0}\makebox(0,0)[b]{\smash{Cluster in Phase Space; $\gamma = 1/16$, $\mu = 1$}}}%
    \put(0.75338491,0.21820425){\color[rgb]{0,0,0}\makebox(0,0)[b]{\smash{Gap in Phase Space; $\gamma = 1/16$, $\mu = 1$}}}%
    \put(0.06865285,0.11139897){\color[rgb]{0,0,0}\makebox(0,0)[b]{\smash{$\rho''$}}}%
    \put(0.24088979,0.00944992){\color[rgb]{0,0,0}\makebox(0,0)[b]{\smash{$\rho$}}}%
    \put(0.46627839,0.03276598){\color[rgb]{0,0,0}\makebox(0,0)[b]{\smash{$\rho'$}}}%
    \put(0.50760051,0.11094257){\color[rgb]{0,0,0}\makebox(0,0)[b]{\smash{$\rho''$}}}%
    \put(0.67983747,0.00899353){\color[rgb]{0,0,0}\makebox(0,0)[b]{\smash{$\rho$}}}%
    \put(0.90522599,0.03230959){\color[rgb]{0,0,0}\makebox(0,0)[b]{\smash{$\rho'$}}}%
  \end{picture}%
\endgroup%